\documentclass[10pt,conference]{IEEEtran} 
\IEEEoverridecommandlockouts

\makeatletter
\newcommand*\titleheader[1]{\gdef\@titleheader{#1}}
\AtBeginDocument{%
  \let\st@red@title\@title
  \def\@title{%
    \bgroup\normalfont\large\centering\@titleheader\par\egroup
    \vskip1.0em\st@red@title}
}
\makeatother

\usepackage{subcaption}
\usepackage{graphicx}
\usepackage[inline,draft,nomargin,index,author=]{fixme}
\usepackage{url}
\usepackage{xcolor}
\usepackage{booktabs}
\usepackage[pdftex,hidelinks]{hyperref}
\usepackage{cleveref}

\crefformat{section}{\S#2#1#3} 
\crefformat{subsection}{\S#2#1#3}
\crefformat{subsubsection}{\S#2#1#3}

\FXRegisterAuthor{nty}{anty}{\textcolor{orange}{Nty}}

\FXRegisterAuthor{kd}{ankd}{\textcolor{red}{Kd}}

\newcommand{\iou}{\textit{io\_uring}}
\newcommand{\fio}{\textit{fio}}
\newcommand{\SPDK}{\textit{SPDK}}
\newcommand{\mq}{\textit{mq-deadline}}

\newcommand{\implicitly}{\textit{implicitly}}
\newcommand{\explicitly}{\textit{explicitly}}
\newcommand{\app}{\texttt{append}}
\newcommand{\rd}{\texttt{read}}
\newcommand{\wrt}{\texttt{write}}
\newcommand{\rst}{\texttt{reset}}
\newcommand{\fns}{\texttt{finish}}
\newcommand{\open}{\texttt{open}}
\newcommand{\close}{\texttt{close}}

\newcommand{\spacemagic}{\vspace{-0.5cm}} 

\usepackage{listings}
\definecolor{mGreen}{rgb}{0,0.6,0}
\definecolor{mGray}{rgb}{0.5,0.5,0.5}
\definecolor{mPurple}{rgb}{0.58,0,0.82}
\definecolor{backgroundColour}{rgb}{0.95,0.95,0.92}

\lstdefinestyle{CStyle}{
    backgroundcolor=\color{backgroundColour},
    emph={Segments,Zones,Extents,sst},
    commentstyle=\color{mGreen},
    keywordstyle=\color{mGreen},
    emphstyle=\color{mGreen},
    numberstyle=\tiny\color{mGray},
    stringstyle=\color{magenta},
    basicstyle=\linespread{1.1}\ttfamily\tiny,
    frame=tlbr,
    framerule=0pt,
    breakatwhitespace=false,
    breaklines=false,
    captionpos=b,
    keepspaces=true,
    numbers=left,
    numbersep=5pt,
    showspaces=false,
    showstringspaces=false,
    showtabs=false,
    tabsize=2
}

\def\BibTeX{{\rm B\kern-.05em{\sc i\kern-.025em b}\kern-.08em
    T\kern-.1667em\lower.7ex\hbox{E}\kern-.125emX}}

\hypersetup{
    colorlinks,
    linkcolor={red!50!black},
    citecolor={blue!50!black},
    urlcolor={blue!80!black}
}

\makeatletter
\renewcommand{\fnum@figure}{Figure~\thefigure}
\makeatother

\makeatletter
\renewcommand{\sectionautorefname}{\S\@gobble}
\renewcommand{\subsectionautorefname}{\S\@gobble}
\makeatother

\newcommand{\figcap}[1]{\caption{\normalfont \textit{#1}}}
\newcommand{\sfigcap}[1]{\caption{\normalfont \textit{\footnotesize #1}}}
\newcommand{\tabcap}[1]{\caption{\normalfont \textit{#1}}}

\newcommand{\obs}[2]{\textbf{Observation \#{#1}}: \textit{#2}~\@}
\newcommand{\rec}[2]{\textbf{Recommendation \#{#1}: \textit{#2}~\@}}

\title{Performance Characterization of NVMe Flash Devices with Zoned Namespaces (ZNS)}

\titleheader{To appear in the \href{https://clustercomp.org/2023/}{IEEE International Conference on Cluster Computing, 2023}. This is an extended report.}

\begin{document}

\author{%
\IEEEauthorblockN{
Krijn Doekemeijer$^{*1}$, Nick Tehrany$^{*1,2}$, Balakrishnan Chandrasekaran$^{1}$, Matias Bj{\o}rling$^{3}$, and Animesh Trivedi$^{1}$}
\IEEEauthorblockA{$^{1}$Vrije Universiteit Amsterdam, Amsterdam, the Netherlands \\
$^{2}$Delft University of Technology, Delft, the Netherlands \\
$^{3}$Western Digital, Copenhagen, Denmark \\
\{k.doekemeijer, n.a.tehrany, b.chandrasekaran, a.trivedi\}@vu.nl, matias.bjorling@wdc.com}
\thanks{*Equal contributions, joint first authors. Nick was with TU Delft during this work.}
}



\maketitle

\thispagestyle{plain}
\pagestyle{plain}

\begin{abstract}
The recent emergence of NVMe flash devices with Zoned Namespace support, ZNS SSDs, represents a significant new advancement in flash storage.
ZNS SSDs introduce a new storage abstraction of append-only \textit{zones} with a set of new I/O (i.e., append) and management (zone state machine transition) commands. 
With the new abstraction and commands, ZNS SSDs offer more control to the host software stack than a non-zoned SSD for flash management, which is known to be complex (because of garbage collection, scheduling, block allocation, parallelism management, overprovisioning).
ZNS SSDs are, consequently, gaining adoption in a variety of applications (e.g., file systems, key-value stores, and databases), particularly latency-sensitive big-data applications.
Despite this enthusiasm, there has yet to be a systematic characterization of ZNS SSD performance with its zoned storage model abstractions and I/O operations.
This work addresses this crucial shortcoming.
We report on the performance features of a commercially available ZNS SSD (13 key observations), 
explain how these features can be incorporated into publicly available state-of-the-art ZNS emulators,
and recommend guidelines for ZNS SSD application developers.
All artifacts (code and data sets) of this study are publicly available at \url{https://github.com/stonet-research/NVMeBenchmarks}.
\end{abstract}


\section{Introduction}
The emergence of fast flash storage in data centers, HPC, and commodity computing has fundamentally effected changes in every layer of the storage stack, 
and led to a series of new developments such as a new host interface (NVM Express, NVMe)~\cite{2022-nvme-2.0,2012-nvme-vs-ahci,2015-systor-nvme-db},
a high-performance block layer~\cite{2013-systor-mq,2019-atc-aync,2019-atc-mqfq,2021-fast-d2fq},
new storage I/O abstractions~\cite{2017-systor-autostream-classification,2014-hotstorage-streamssd,2014-asplos-sdf,2021-atc-zns,2021-iouring1,2021-iouring2,2022-systor-iouring},
and re/co-design of storage application stacks~\cite{2018-tos-flashnet,2016-hotstorage-nvmedirect,2021-osdi-blkswitch,2017-tos-iris,2021-osdi-int-cali,2023-arxiv-fs-survey,2022-arxiv-kv-survey}. 
Today, flash-based solid-state drives (SSDs) can support very low latencies (i.e., a few microseconds), and multi GiB/s bandwidth with millions of I/O operations per second~\cite{2019-vault-storage-perf,2021-spdk-perf,2021-axboe-10m}.



Despite these advancements, the conceptual model of a storage device remains unchanged since the introduction of hard disk drives (HDDs) 
more than half a century ago.  
A storage device supports only two necessary operations: write and read data in units of \textit{sectors} (or blocks)~\cite{2022-ostep}.
Data can be read from and written to anywhere on the device, hence supporting random and sequential I/O operations.  
Though this model works with conventional HDDs, it is not apt for flash-based storage devices as flash internally does not support overwriting data~\cite{2017-eurosys-unwritten,2022-systor-ssd-study,2008-atc-ssd-tradeoff}. 
Flash devices offer the illusion of ``overwritable'' storage via the \textit{flash translation layer (FTL)}, a software component that runs within the device.
The FTL enables easy integration of flash devices (by allowing them to masquerade as fast HDDs), albeit it introduces unpredictability in performance~\cite{2009-cidr-uflip,2020-osdi-linnos,2017-fast-tiny-tail,2019-atc-swan,2021-tos-ssd-blackbox,2013-sigmetrics-ssd-expectations} and complicates device lifetime management~\cite{2021-sigmetric-ssd-management}. 
These challenges are defined as the \textit{unwritten contracts} of SSDs~\cite{2017-eurosys-unwritten}. 
As data centers have largely transitioned to SSDs for fast, reliable storage~\cite{andersen2010rethinking,han2021depth}, and modern big data applications have high QoS demands~\cite{dean2013tail, balmau2019silk}, there is a dire need to address these unwritten contracts.

Researchers and practitioners advocate for open flash SSD interfaces beyond block I/O~\cite{2013-cidr-death-block} to address these challenges.
%
Examples include 
\textit{Open-Channel SSDs} (OCSSD)~\cite{2017-fast-openchannel-ssd}, \textit{multi-stream SSDs}~\cite{2014-hotstorage-streamssd}, and, more recently, \textit{Zoned Namespaces} (ZNS)~\cite{2021-atc-zns}. 
The focus of this work is on NVMe devices that support ZNS, which are commercially available today~\cite{2021-samsung-zns,2021-wd-zns}.
ZNS promises a low and stable tail latency~\cite{2021-atc-zns} and a high device longevity, and, hence, addresses the needs of modern big-data workloads.
There is, unsurprisingly, a rich body of active and recent work on ZNS~\cite{2020-nvmsa-zns-implications,2021-atc-zns,2020-vault-append,2021-hotos-zone,2022-cidr-append,2021-hotstorage-rocks,2021-atc-znswap,2022-hotstorage-zns-parallelism,2022-hotstorage-comp-lsm,2022-zenfs,2019-fast-geardb,2022-hotstorage-llc,2021-osdi-zns+,oh2023zenfs+}.
%
Despite this enthusiasm, there has \textit{not} been a systematic performance and operational characterization of ZNS SSDs.
%
%
The lack of an extensive characterization of ZNS SSDs severely limits the utilization and application of ZNS devices in big-data workloads.
In this work, we bridge this gap by presenting the performance characterization of a commercially-available NVMe ZNS device.


We complement this characterization of a physical device with an investigation of emulated ZNS devices,
since they are widely used in research~\cite{2022-hotstorage-comp-lsm, liu2022fair, oh2021efficient, 2021-osdi-zns+}.
Emulated devices enable researchers to explore the ZNS design space without being constrained by device-specific characteristics.
Such unconstrained explorations are crucial since ZNS is a new interface and the selection of available configurations in a real SSD is, unsurprisingly, quite limited.
%
%
The research validity of all of these works hinge on an emulator's ability to mimic the performance characteristics of real hardware.
In our investigation of the state-of-the-art emulators---FEMU~\cite{li2018case} and NVMeVirt~\cite{kim2023nvmevirt}---we reveal limitations in ZNS performance characterizations and discuss approaches to address them.

We summarize our key contributions as follows.

$\bullet$~We systematically characterize the performance (i.e., latency, bandwidth, and parallelism management) of a commercially available NVMe ZNS device, including its \app{}, \rd{}, \wrt{}, and zone management operations (i.e., \fns{} and \rst{}).

$\bullet$~We analyze the implications of interference from ZNS's unique \rst{} and \app{} operations on the \rd{}, \wrt{} and \app{} I/O performance.

$\bullet$~We reveal limitations in the performance models of the state-of-the-art ZNS emulators and discuss recommendations to address them.

$\bullet$~We share key recommendations for ZNS application developers using the insights from our characterizations.

$\bullet$~We publish our benchmarking software and data set at \url{https://github.com/stonet-research/NVMeBenchmarks} for encouraging reproducible research.

\section{Background}\label{sec:background}
In this section, we review the background on flash-based storage and ZNS SSDs.

\subsection{Flash storage}

The storage area of flash-based devices is organized into flash \textit{pages}, which is the unit of addressing and I/O operations~\cite{2012-cacm-ssd-anatomy}.
A typical flash page is 4--16\,KiB, which is always atomically written or programmed~\cite{2022-systor-ssd-study}.
%

A flash page can not be overwritten; rather the page \textit{must} be erased before it can be written again.
Flash SSDs, nevertheless, provide an illusion of over-writable storage by storing data (i.e., the overwrite) in a new (flash) page, while marking the old (i.e., replaced) page as invalid or ``garbage.''
After these garbage pages are eventually erased, the pages can be (re)used for writing data.

Erasing a page is a complex and time-consuming operation on flash storage.
First, pages are physically grouped into \textit{blocks}, and erasures only work at the block level.
Second, since a block may contain both valid and garbage pages, valid pages have to be copied to a new block prior to erasing the old block.
Third, pages always need to be sequentially written to blocks, complicating data management as random writes are not allowed.
The erasing of a block and the associated book-keeping tasks (e.g., tracking valid and garbage pages) collectively constitute the \textit{garbage collection (GC)} process (Flash-based SSDs, Chapter-44,~\cite{2022-ostep}).
Such GC processes may interfere with concurrent I/O operations and degrade I/O performance, since the GC process and user-issued I/O operations may need to access the same block or storage area~\cite{2014-ieee-gc-past-present-future}.
Further on, flash memory has limited lifetime because it has a \textit{limited} write or program/erase (P/E~\cite{yamada1993degradation}) \textit{endurance}. It is also prone to \textit{read-write disturbs} that occur when many reads are issued to the same blocks---they cause data loss.
Limited storage lifetime exacerbates GC and other management (e.g., lifetime optimization) tasks~\cite{2022-systor-ssd-study, 2012-cai-error, 2013-cai-threshold, 2015-cai-data, 2015-cai-read, 2015-systor-read-interferences}.

The onus of hiding these complexities of flash SSD management from applications running on the host resides with the \textit{flash translation layer} (FTL), which is a part of the device firmware.
It provides, for instance, the familiar page-addressable, any-address readable or writable storage media~\cite{2014-csur-ftl-translation,2022-ostep}.
Several prior work offer in-depth explanations on flash drives and FTLs~\cite{bez2003introduction, micheloni2010inside, aritome2015nand,2014-ieee-gc-past-present-future,2022-systor-ssd-study}.



\subsection{Devices with Zoned Namespace}\label{sec:background_zns}

\begin{figure}[t!]
\centering
  \includegraphics[width=0.9\linewidth]{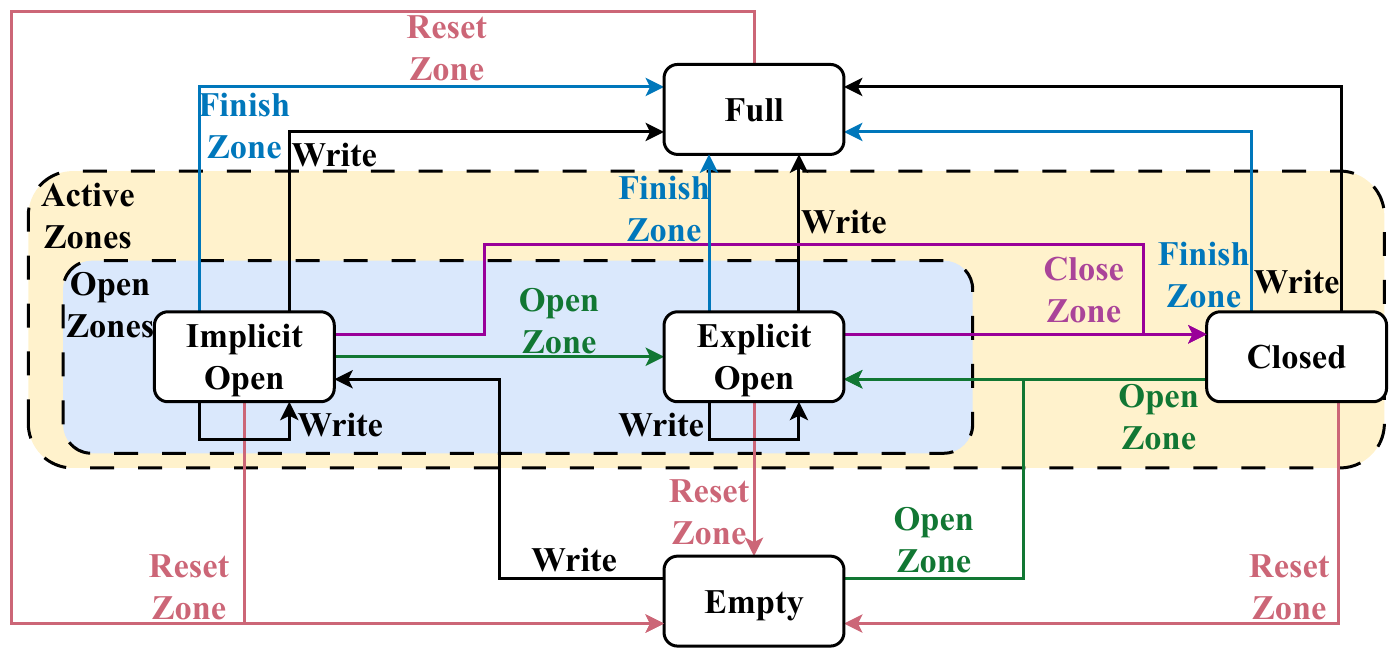}
  \figcap{Simplified overview of the ZNS zone state machine}\label{fig:zns-state-diagram}
\spacemagic{}
\end{figure}

The storage area in ZNS devices is divided into regions known as \textit{zones}. 
Zones themselves are divided into blocks, which represent the fundamental unit of I/O operations---the blocks are logically equivalent to pages in traditional flash storage.
%
Blocks are addressed using a \textit{logical block address (LBA)}.
%
Zones only support sequential writes, matching the constraint of underlying flash pages, which can only be programmed sequentially.
%
To meet this sequential-write constraint, within a zone, applications must issue page-write requests to LBAs in increasing order.
SSDs, however, schedule I/O requests internally and may reorder them as required~\cite{2021-fast-d2fq}.
There can, hence, only be a single in-flight \wrt{} request per zone so that the re-orderings, if any, do not violate the must-be-sequentially-written constraint of a zone.
This design, consequently, limits \wrt{} concurrency in a zone to one.

ZNS introduces the \app{} operation~\cite{2020-vault-append} to alleviate the \wrt{} concurrency limitation within a zone.
The \app{} idea is similar to \textit{nameless writes}~\cite{2012-fast-nameless}, writes that are not issued to an address, but return the address on completion.
Unlike the \wrt{} operation, which accepts the target block address, the \app{} operation takes a \textit{zone starting} LBA (ZSLBA) along with the data. 
Once an \app{} is completed, the LBA is returned to the application. 
As a result it is safe to reorder \app{}s in a zone, which enables applications to issue multiple outstanding \app{} requests to the same zone.

ZNS SSDs offer explicit control over their GC process through the \rst{} operation.
A \rst{} operation on a zone informs the device that the data in the zone can be discarded and the zone can be garbage collected.
The \rst{} operation does not, nevertheless, immediately force a block erasure~\cite{jung2023preemptive}.
The \rst{} operation can be limited to FTL-metadata-mapping manipulations, which indicate to the device that the block can be erased later.


Zones of a ZNS device have states (Fig.~\ref{fig:zns-state-diagram}), which dictate the allowed operations on a specific zone.
Since each zone operation (e.g., \rd{}, \wrt{}, and \app{}) consumes SSD resources (e.g., internal buffers), there are limits on the number of zones that can be concurrently opened and used.
These limits are defined as the \textit{maximum open zone limit} and \textit{maximum active zone limit}, respectively.
Applications must abide by these constraints, and explicitly manage the zone states and transitions.
An application must, for instance, open a zone \textit{before} it accepts \wrt{}s or \app{}s.
State transitions can be internal to a device and \textit{implicit} (e.g., a \wrt{} to an empty zone transitions it to an open zone in Fig.~\ref{fig:zns-state-diagram}), or \textit{explicit} as a response to a user request.

ZNS offers several explicit zone management operations, which include \open{}, \close{} and \fns{}.
We skip discussing the first two, whose names reveal their functionalities, and focus on the last.
The \fns{} operation transforms an \textit{open} zone directly into a \textit{full} zone.
It releases all resources attached to the zone (to stay within the maximum open zone limit).
Then, the device can either fill the zone with data or mark the unused capacity with mapping (metadata) updates in the ``finished'' zone (Fig.~\ref{fig:zns-state-diagram}).
Mapping updates would require extra metadata to keep track of partially-filled zones.
The \fns{} operation has implications for performance, and the costs of this operation varies from one ZNS SSD implementation to another.

In summary, ZNS devices support a rich I/O interface that includes operations beyond the simple \rd{} and \wrt{} operations of traditional flash storage.
It is, therefore, crucial to understand and characterize the performance of these operations as they (and their state-machine transitions) are now part of the Linux storage software stack.



\subsection{Software support}

ZNS devices are fully supported in Linux since kernel version \texttt{5.9}~\cite{2022-zns-intro}. 
Currently there is a limited number of applications that use ZNS, and most that do, do not use all functionalities (e.g., no \fns{} or \open{}).
Evaluating these applications would limit what ZNS properties we can measure and, therefore, in our work we use synthetic benchmarks to understand all of ZNS' facets first.
The results of our benchmarks can then be used for application design.
Here, we briefly mention several prominent ZNS applications in research to present an overview of what is currently available.
Currently, applications have access to ZNS-friendly file systems F2FS~\cite{2015-fast-f2fs}, Btrfs~\cite{rodeh2013btrfs} and Ceph~\cite{ha2023zceph}.
There is also support for a swap system known as ZNSwap~\cite{2021-atc-znswap} and a RAID system known as ZRAID~\cite{kim2023raizn}
Lastly, KV-store RocksDB has ZenFS as a ZNS-capable file system back-end~\cite{2021-atc-zns}.

\begin{table}[t]
\def\arraystretch{1.25}
\footnotesize
\centering
\tabcap{Overview of the key insights}\label{tab:key-findings}
\begin{tabular}{p{0.3\linewidth}| p{0.6\linewidth}}
\toprule
\textbf{\textit{Category}} & \textbf{\textit{Insight}}\\
\midrule
\textit{\texttt{Append} vs. \wrt{}} &  \texttt{Write} operations have up to 23\% lower latencies than \app{} operations (\S\ref{sec:exp-1-app-write-lat})\\
\textit{Scalability} & Prefer intra-zone scalability (\S\ref{sec:exp-2-scalability},~\S\ref{sec:exp-3-sm})\\
\textit{Zone transitions} & \texttt{Finish} is the most expensive operation; it takes up to hundreds of milliseconds (\S\ref{sec:exp-3-sm})\\ 
\textit{I/O interference} &  NVMe ZNS offers 3$\times$ higher read throughput under concurrent I/O operations than NVMe (\S\ref{sec:exp-4-inf-io}) \\  
\textit{I/O \& GC interference} & \texttt{Reset} latency increases by up to 78\% under concurrent I/O operations, but \rst{} operations themselves have no effect on \app{}, \rd{} or \wrt{} operations (\S\ref{sec:exp-5-inf-reset})\\ 
\bottomrule
\end{tabular}
\spacemagic{}
\end{table}

\section{Experiments}

In this paper, we characterize the performance and interference properties of the
\href{https://www.westerndigital.com/products/internal-drives/data-center-drives/ultrastar-dc-zn540-nvme-ssd#0TS2094}{ Western Digital Ultrastar DC ZN540} SSD, a large-zone ZNS SSD, using a series of controlled benchmarks. 
As of this writing, the number of commercially-available ZNS SSDs is limited, therefore, we focus our efforts on characterizing one SSD model and synthesize the performance questions to ask when evaluating a ZNS SSD.
\autoref{tab:key-findings} summarizes our key findings.

\subsection{Benchmarking setup}\label{sec:setup}

We use \fio{}~\cite{2021-fio} for generating the workloads and benchmarking the ZNS device.
We also employ custom \SPDK{} benchmarks for benchmarking state transitions (\cref{sec:exp-3-sm}) and \rst{} interference (\cref{sec:exp-5-inf-reset}), since \fio{} does not support them.
We describe our benchmarking platform in detail in~\autoref{tab:benchmarking-setup}.


We use two storage stacks for benchmarking: the Linux kernel block layer and the \SPDK{} stack.
%
%
The Linux block layer ships with the \textit{mq-deadline} scheduler, which buffers multiple \wrt{} operations to a single zone, merges writes to contiguous LBAs into one or multiple (larger) writes, and sequentially issues the merged requests.
Applications can, hence, issue multiple \wrt{} operations to a single zone.
The \SPDK{} stack, in contrast, is a bare-bones storage stack without any I/O scheduler.
The rationale behind our storage stack selection is twofold.
First, no storage stack currently supports all combinations of I/O and management operations that we aim to benchmark.
We cannot, for instance, issue and benchmark \app{} or zone management operations via \fio{} and the Linux I/O stack.
In a similar vein, we are restricted to issuing only one \wrt{} per zone at a time with \SPDK{}, since it lacks an I/O scheduler.
Second, the selection enables us to compare the implications of state-of-the-practice---the Linux stack---and that of the state-of-the-art---\SPDK{}---for ZNS application development.
%

We run experiments for 20 minutes and/or repeat them at least three times for deriving robust statistics.
We pin our benchmarking code to the NUMA node containing the ZNS device.
For the Linux storage stack, we use the \iou{} engine with submission-queue polling enabled, following the recommended settings~\cite{2022-systor-iouring}.

\begin{table}[t!]
\footnotesize
\def\arraystretch{1.25}
\centering
\tabcap{Details of the benchmarking environment}\label{tab:benchmarking-setup}
\begin{tabular}{p{0.15\linewidth}| p{0.75\linewidth} }
\toprule
\textbf{\textit{Component}} & \textbf{\textit{Configuration details}}\\
\midrule
CPU & Dual socket Intel(R) Xeon(R) Silver 4210 CPU @ 2.20GHz, 2 sockets, 10 cores/socket, hyper-threading disabled, with Spectre and Meltdown patches, \texttt{intel\_pstate=disable, intel\_idle.max\_cstate=1}\\
DRAM & 256 GiB, DDR4\\
ZNS & \href{https://www.westerndigital.com/products/internal-drives/data-center-drives/ultrastar-dc-zn540-nvme-ssd#0TS2094}{Western Digital Ultrastar DC ZN540 1TB}\newline(zone size: 2,048~MiB; zone capacity: 1,077~MiB; total number of zones: 904; max. active zones: 14)\\
NVMe &  \href{https://www.westerndigital.com/products/internal-drives/data-center-drives/ultrastar-dc-sn640-nvme-ssd#0TS1849}{Western Digital Ultrastar DC  SN640 960 GB}\\
Software & Ubuntu (v22.04), kernel (v5.19, built from source), fio (v3.32; \href{https://github.com/axboe/fio/tree/db7fc8d864dc4fb607a0379333a0db60431bd649}{\#db7fc8d}), SPDK (v22.09; \href{https://github.com/spdk/spdk/tree/aed4ece93c659195d4b56399a181f41e00a7a25e}{\#aed4ece}), nvme-cli tools (v2.0, \href{https://github.com/linux-nvme/nvme-cli/tree/5a36baba322e7979175ed9dfeede0a96e29efc37}{\#5a36bab})
    \\
\bottomrule
\end{tabular}
\spacemagic{}
\end{table}

\subsection{Performance metrics}

We briefly describe the metrics we use to evaluate the performance of NVMe (ZNS) devices. 
Two indicators of I/O operation performance are throughput (i.e., the number of operations or bytes per second) and operation latency (i.e, the time each operation takes). 
We measure ZNS throughput in I/O operations completed per second, referred to as IOPS, or in bytes written/read per second.
We measure operation latency from the moment a request is submitted on the NVMe submission queue until a request is completed and visible on the NVMe completion queue.
It is possible to send requests at a higher \textit{queue depth} (QD)---QD measures the number of requests that can be concurrently in flight. 
When the queue depth is higher than 1, it is possible that multiple requests are submitted, but not yet in a completed state.
This has implications for request latency as some requests will take longer to complete than others. We always, hence, mention the queue depth of an experiment.

%
%

\subsection{\app{} and \wrt{} performance}\label{sec:exp-1-app-write-lat}

%
\app{} and \wrt{} operations both write data to the device, albeit they differ in their approach (refer~\autoref{sec:background_zns}).
The difference between \app{} and \wrt{} operations, fundamentally, lies in who is responsible for ensuring sequential writes to a zone---host (in case of a \wrt{}) or device (in case of an \app{}).
Currently, there is no standard benchmark to make an educated decision on what operation to use.
We perform, therefore, a quantitative analysis of the performance of both operations and facilitate making an informed decision about the use of these operations, taming the complexity and determining the changes required in the storage stack (see file system design for nameless writes~\cite{2012-fast-nameless}). 

We evaluate \wrt{} and \app{} operations as follows.
First, we study them under varying LBA formats (format of the NVMe namespace) with sector sizes 512\,B and 4\,KiB to verify whether \app{} and \wrt{} operations are both affected by the format.
We then select the LBA format that results in the lowest latency.
Second, we investigate the implications of the choice of I/O engine (i.e., \iou{} and \SPDK{}) and scheduler (i.e., \textit{none} and \textit{mq-deadline}) on the performance of these operations.
This investigation helps to decide what engine to use for ZNS and whether to use a host scheduler or appends.
Third, we evaluate the effect of request sizes on I/O latency.
%
The evaluations are single-threaded and synchronous (QD=1) in order to evaluate the performance of requests in isolation. 
While prior work demonstrates that request size affects \wrt{} latency on ZNS~\cite{2020-nvmsa-zns-implications}, we investigate if this observation is also apparent for \app{} latency.
%

\begin{figure}[t]
    \centering
    \begin{subfigure}[b]{0.45\linewidth}
        \includegraphics[width=1\linewidth]{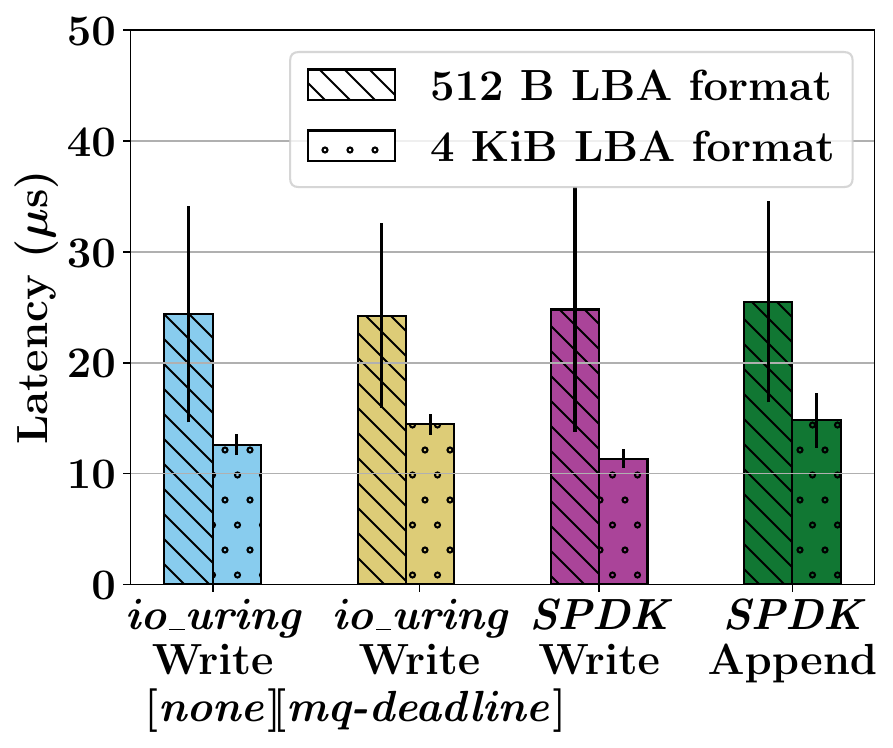}
        \sfigcap{Request sizes:\\ 512\,B for 512\,B LBA format;\\ 4\,KiB for 4\,KiB LBA format}
        \label{fig:f1-comparison}
    \end{subfigure}%
    \begin{subfigure}[b]{0.45\linewidth}
        \includegraphics[width=1\linewidth]{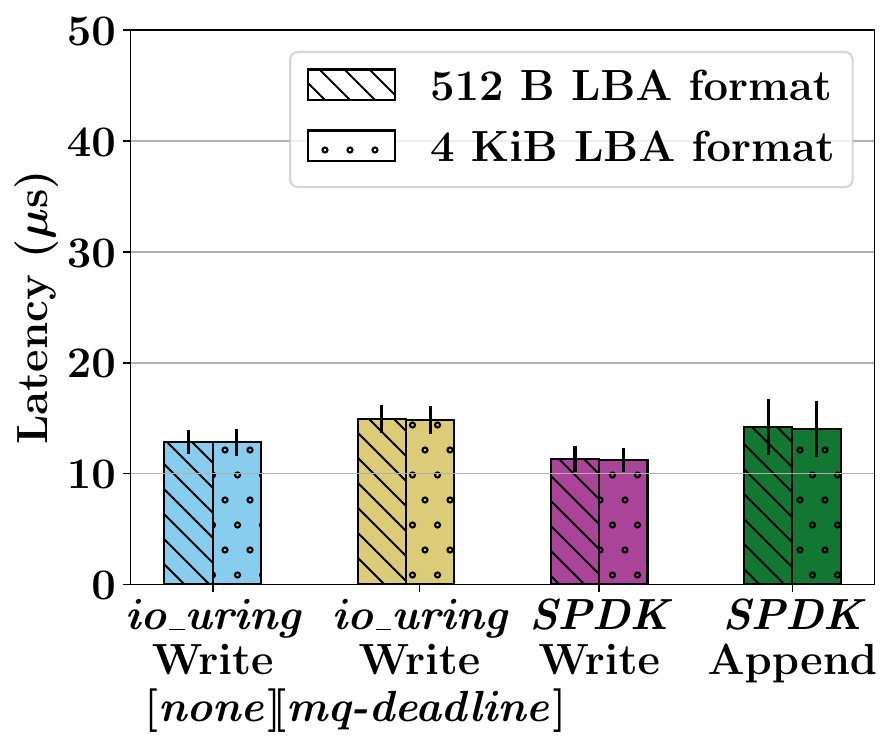}
        \sfigcap{Request sizes:\\ 4\,KiB for \wrt{}s;\\ 8\,KiB for \app{}s}
        \label{fig:f1-optimal-req-size}
    \end{subfigure}
    \figcap{I/O latencies of \app{} and \wrt{} operations (queue depth, QD=1)}\label{fig:f1}
\end{figure}

\obs{1}{The LBA format can have significant impact on both \wrt{} and \app{} latencies.}
The LBA format affects both \app{} and \wrt{} operations~(\autoref{fig:f1-comparison}).
We show the operational latencies in microseconds along the Y-axis (lower is better) and software stack combinations with 512\,B and 4\,KiB formats along the X-axis.
We set the request size to the same value as the block size of the respective LBA format.
Latencies of the operations with a 4\,KiB LBA consistently outperform that of a 512\,B LBA, sometimes by as much as a factor of two (\autoref{fig:f1-comparison}).
%
This difference between formats is highly dependent on the firmware, as firmware might not be optimised for small I/O, but highlights that it is important to consider what format to use for applications/benchmarks.
%
%
Before running experiments we pick the optimal format size.
We, hence, use the 4\,KiB LBA format for all further experiments.
We note, however, that the choice of an optimal block size depends on the ZNS device: It may be that the device was, for instance, explicitly optimized for 512\,B accesses.
We, therefore, need to consider what LBA format to use for both \wrt{} and \app{} operations.

%

\obs{2}{Using the \SPDK{} storage stack results in the lowest latencies.}
The Linux storage stack has, typically, higher overheads than the \SPDK{} stack~\cite{didona2022understanding}, and we report that it holds true for ZNS as well (see the 4\,KiB format in~\autoref{fig:f1-comparison}).
\SPDK{} has the lowest latency overheads for single outstanding I/O requests:
\wrt{} operations in \SPDK{} (11.36\,$\mu$s) have 9.98\% lower latency than that in the kernel without a scheduler (12.62\,$\mu$s).
The \textit{mq-deadline} scheduler, furthermore, adds non-negligible latency overheads (i.e., 1.85\,$\mu$s out of 14.47\,$\mu$s, or 12.81\%).
As the latency of raw flash storage access decreases, the scheduler's relative overheads will increase~\cite{2023-cheops-performance-characterization}.

\begin{figure}[tb]
    \centering
    \begin{subfigure}[b]{0.45\linewidth}
        \includegraphics[width=1\linewidth]{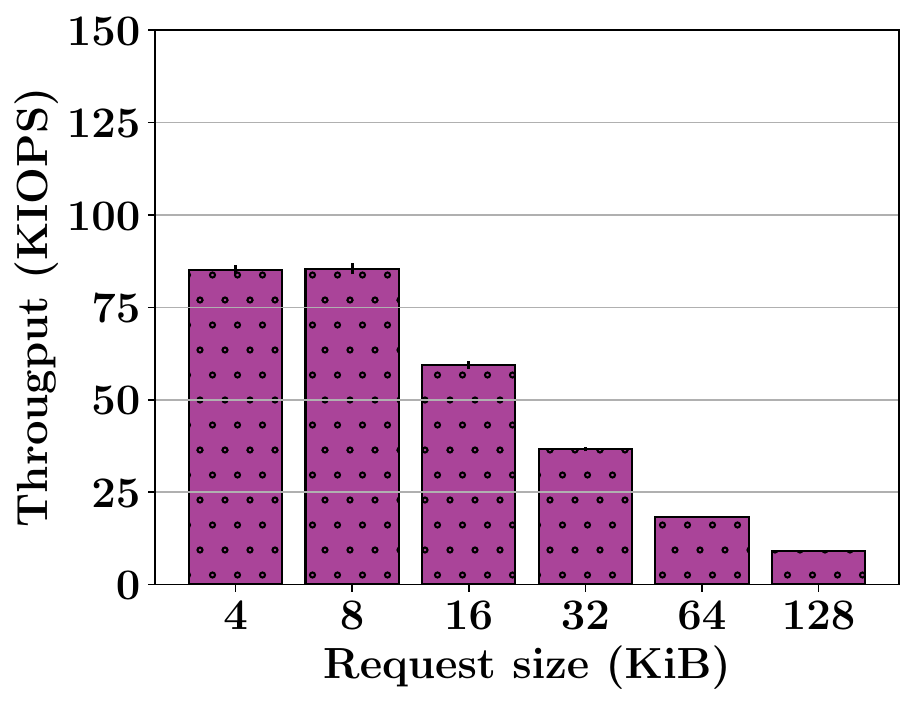}
        \sfigcap{\wrt{}}
        \label{fig:spdk-write-req-size}
    \end{subfigure}%
    \begin{subfigure}[b]{0.45\linewidth}
        \includegraphics[width=1\linewidth]{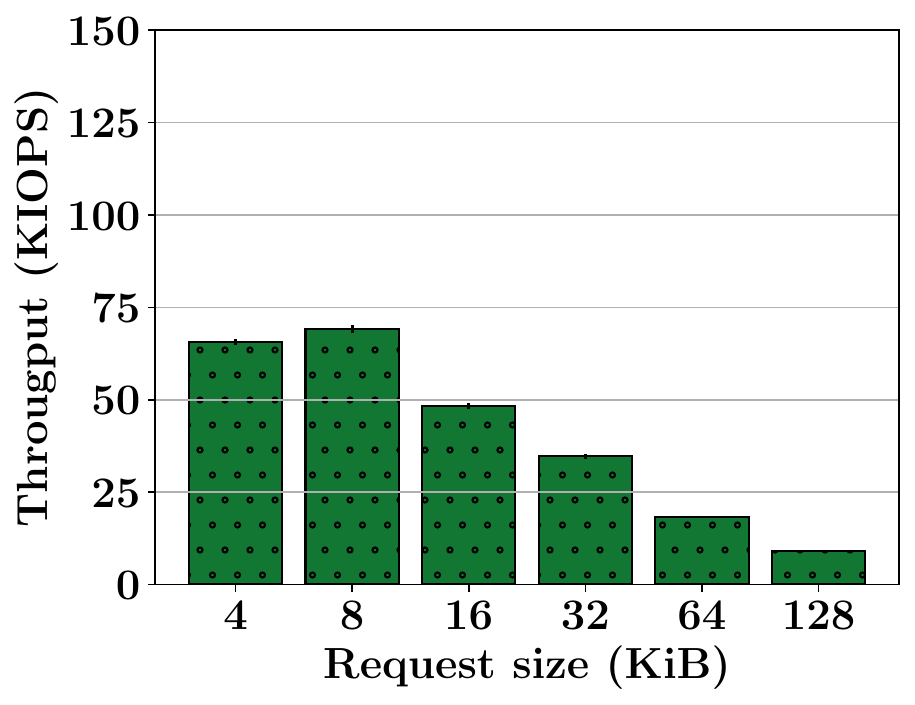}
        \sfigcap{\app{}}
        \label{fig:spdk-append-req-size}
    \end{subfigure}
    \figcap{SPDK I/O throughput in IOPS (QD=1)}
    \label{fig:f2}
\end{figure}

\obs{3}{\texttt{Write} and \app{} throughput depends on the request size.}
We plot the throughput (in terms of thousands of I/O operations per second) of \wrt{} (in \autoref{fig:spdk-write-req-size}) and \app{} operations (in \autoref{fig:spdk-append-req-size}) as a function of request size (in KiB).
We observe that \wrt{} operations experience the highest throughput in IOPS (i.e., 85\,KIOPS) for request sizes of 4\,KiB and 8\,KiB, whereas the performance of \app{} operations improves slightly---from 66 to 69\,KIOPS---when we double the request size.
The throughput in bytes per seconds is highest for large requests (e.g. $>=$32\,KiB, calculated as request size $\times$ IOPS).
Note that we issue requests synchronously, hence, throughput is the inverse of request latency.
%
The impact of request size on performance, hence, differs between \app{} and \wrt{} operations.
That the \app{} throughput is lower than \wrt{} throughput is not inherent to the design of ZNS and dependent on the firmware.
It is expected that \app{} throughput will increase for newer ZNS devices.
It is likely that the request size has an impact because of zone parallelism (i.e., zones mapped to multiple flash channels), similar to what is observed in \cite{2020-nvmsa-zns-implications}.
We, therefore, recommend issuing large requests for maximal throughput;
for the ZNS device used in our evaluation, we observe maximal throughput in IOPS with 4\,KiB and 8\,KiB for \wrt{} and \app{} operations, respectively, and maximum throughput in bytes for requests larger than 32\,KiB.

\begin{figure*}[tbp]
  \centering
  \begin{subfigure}[t]{0.3\linewidth}
    \includegraphics[width=\linewidth]{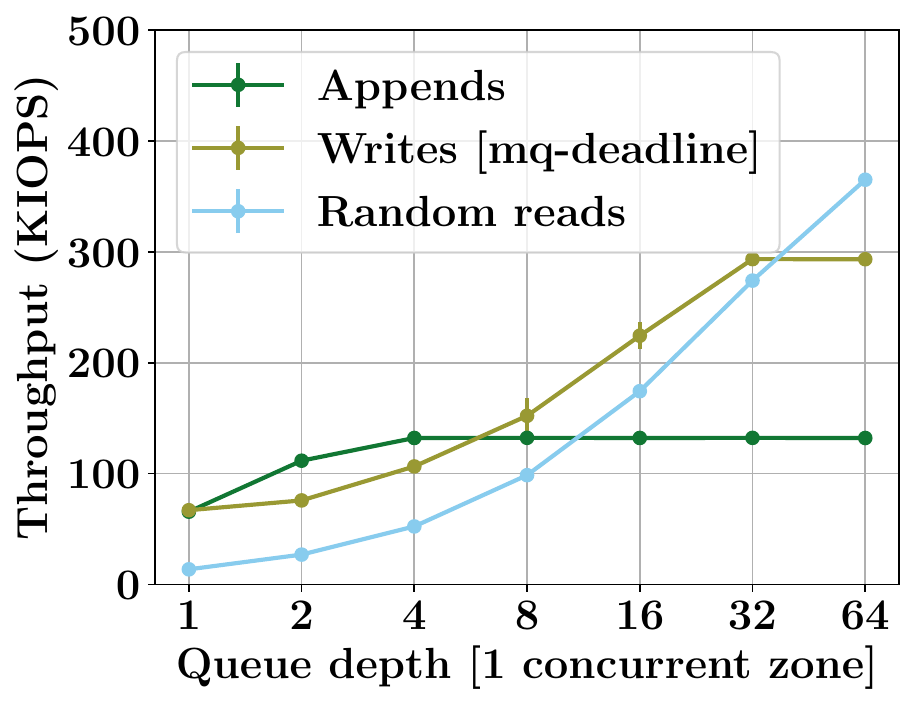}
    \sfigcap{}\label{fig:intra_throughput}
  \end{subfigure}
  \begin{subfigure}[t]{0.3\linewidth}
    \includegraphics[width=\linewidth]{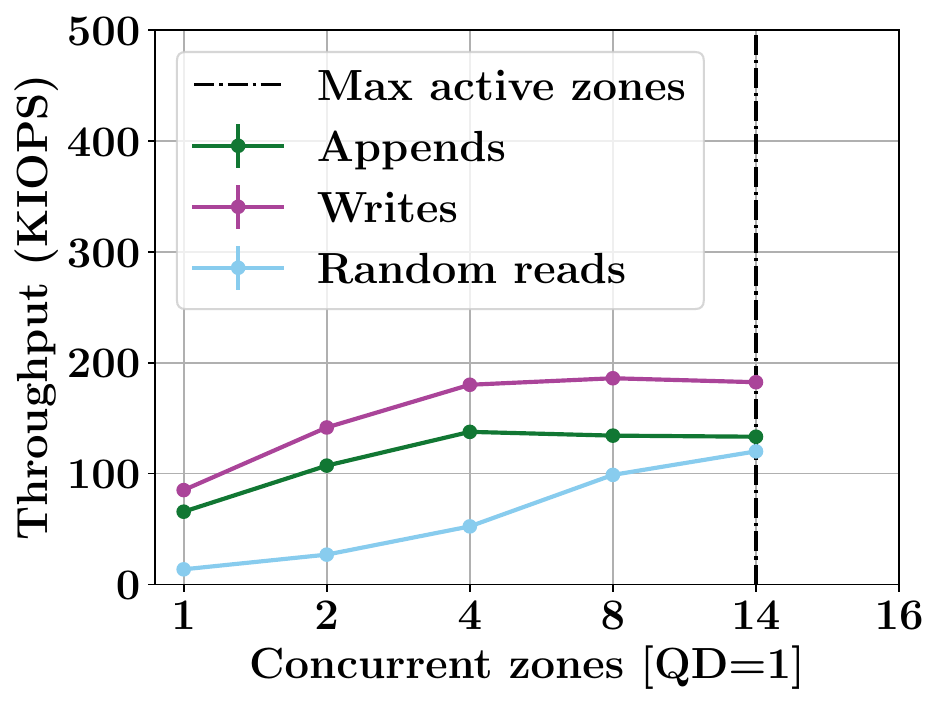}
    \sfigcap{}\label{fig:inter_throughput}
  \end{subfigure}
  \begin{subfigure}[t]{0.3\linewidth}
    \includegraphics[width=\linewidth]{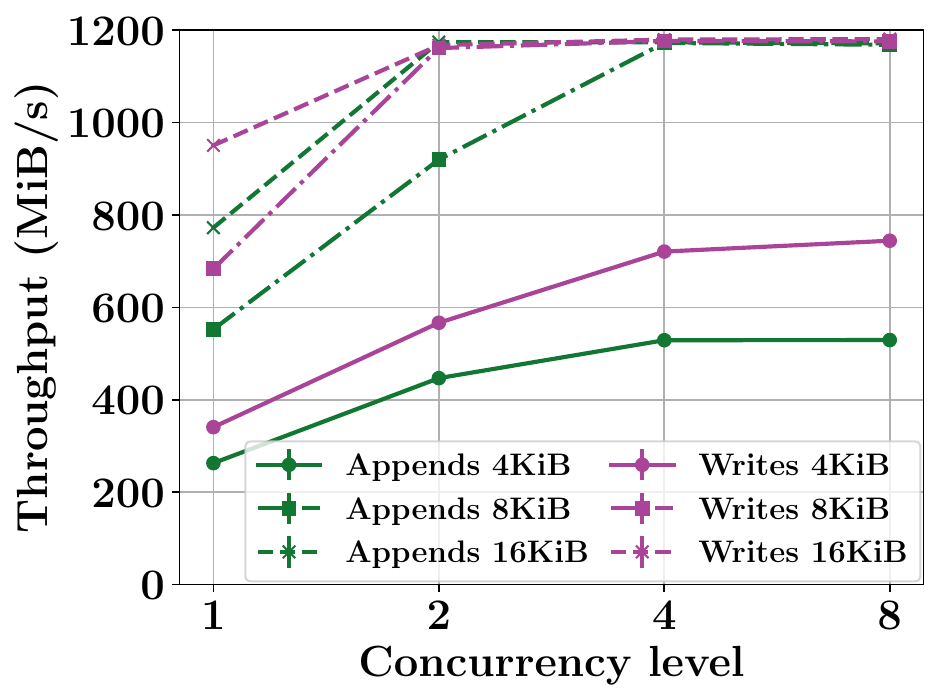}
    \sfigcap{}\label{fig:inter-versus-intra-bandwidth}
  \end{subfigure}
  \figcap{(a) Intra-zone scalability in IOPS for 4\,KiB requests (variable QD, 1 zone/thread); (b) Inter-zone scalability in IOPS for 4\,KiB requests (QD 1, variable zones/threads); and (c) Inter-zone and intra-zone bandwidths using \SPDK{}; concurrency level is queue depth for \app{} operations, and concurrent zones for \wrt{}s}
\end{figure*}

\obs{4}{\wrt{} operations have lower I/O latencies than \app{} operations.}
Across all configurations (\autoref{fig:f1-comparison}, \autoref{fig:f1-optimal-req-size}, \autoref{fig:spdk-write-req-size}, \autoref{fig:spdk-append-req-size}), we observe that the latency of \wrt{} operations is lower than that of \app{} operations, even if the request size is the same.
In \autoref{fig:f1-optimal-req-size}, we use  4\,KiB \wrt{} and 8\,KiB \app{} operations, since these request sizes offered the lowest I/O latencies in prior experiments, and retain them as such for both the 512\,B and 4\,KiB block-size LBA formats.
Our request sizes are now multiples of the block size, and they show, hence, fewer overheads.
We achieve low \wrt{} latencies of 11.36\,$\mu$s (4\,KB with \SPDK{} \wrt{}) and \app{} latencies of 14.02\,$\mu$s (8\,KiB with \SPDK{} \app{}), and observe differences as large as 3.48\,$\mu$s (or 23.42\%) between \wrt{} and \app{} operations.

\rec{1}{Use \wrt{} instead of \app{} operations for low I/O latencies (differences between them can be as much as 23\%), and use the SPDK storage stack since it delivers the lowest I/O latencies.}

\subsection{Scalability: intra-zone versus inter-zone}\label{sec:exp-2-scalability}


A set of I/O requests such as \rd{}, \wrt{}, and \app{} operations can be distributed over either a single zone (intra-zone scalability) or multiple zones (inter-zone scalability).
We define the maximum number of in-flight requests for both intra- and inter-zone as the \textit{concurrency level}.
Below, we analyze both of these approaches.


We measure the scalability of random \rd{}, sequential \wrt{}, and sequential \app{} operations in terms of IOPS and bandwidth.
In this setup, we always issue \rd{} and \app{} operations via \SPDK{}, but we issue \wrt{} operations via \SPDK{} for measuring inter-zone scalability, and use Linux with \textit{mq-deadline} and \iou{} for measuring intra-zone scalability.
We use \SPDK{} wherever possible in the evaluations owing to its low overhead (\textit{Observation~\#2}) and \app{} support.
We rely on \iou{} for issuing multiple \wrt{} operations to the same zone, since \SPDK{} (with no access to schedulers) does not support this functionality.
Intra-zone benchmarks are single-threaded and inter-zone benchmarks use one thread for each concurrent zone.

We measure the throughput of sequential \wrt{}, \app{}, and random \rd{} operations using 4\,KiB requests both within a single (\autoref{fig:intra_throughput}) and across multiple zones (\autoref{fig:inter_throughput}).
We plot the throughput (in KIOPS) as a function of the level of concurrency in the system---in terms of queue depth for intra-zone and zones for inter-zone scalability measurements.
Below we discuss three primary observations; all of them reveal that intra-zone parallelism fares better than inter-zone parallelism.



\obs{5}{Intra-zone parallelism achieve higher overall IOPS than inter-zone parallelism.}
We observe that both \rd{} and \wrt{} intra-zone operations (\autoref{fig:intra_throughput}) achieve higher IOPS than inter-zone requests (\autoref{fig:inter_throughput}) as we increase the level of concurrency.\footnote{The \wrt{} operations in \autoref{fig:intra_throughput} use \iou{} with the \texttt{mq-deadline} scheduler, while the \wrt{} operations in \autoref{fig:inter_throughput} use \SPDK{}.}
We also note that inter-zone scalability is further constrained by the maximum
open zone limit (\autoref{sec:background_zns}):
The number of concurrent zones when issuing \wrt{} or \app{} operations to multiple zones were, for instance, limited to a maximum of 14 zones, which was the maximum open zone limit for the device we evaluated.
We prefer intra-zone to inter-zone parallelism if applications require higher scalability than permitted by the number of open zones.


\obs{6}{The \app{} throughput, however, is agnostic to whether we use inter-zone or intra-zone requests.}
Either scaling method offers similar throughput for \app{} operations.
The \app{} throughput (in \autoref{fig:intra_throughput} and \autoref{fig:inter_throughput}) increases slightly until concurrency level 4 ($\sim$132\,KIOPS), but does not improve afterwards.
We hypothesize this limit is the device limit---not fundamental to the implementation or design of the ZNS device.
That the \wrt{} operations exhibit only marginal increase, if any, in throughput beyond 4 concurrent zones supports our hypothesis.
The results indicate that distributing \app{} operations across zones is a valid scaling strategy too, albeit it wastes additional open zones.


\obs{7}{In a single zone \rd{} operations scale the best (with queue depth), followed by \wrt{} and \app{} operations.}
The \app{} throughput reaches a maximum of 132\,KIOPS at a queue depth of 4.
The throughput of \wrt{} operations with \textit{mq-deadline}, in contrast, reaches 293\,KIOPS at a queue depth of 32.
However, the \rd{} throughput, reaches 424\,KIOPS at a queue depth of 128 (although not shown in the figure).
Until a queue depth of 4, within a single zone, \app{} operations outperforms \wrt{} operations, but at higher queue depths \wrt{} operation outperforms \app{} operations.
This drastic performance improvement of \wrt{} operations stems from the \textit{mq-deadline} scheduler:
The \wrt{} operations to a zone must be issued sequentially, and the scheduler merges these sequential (4\,KiB requests) into (fewer) larger \wrt{} operations, thereby delivering higher throughput than those of the isolated \wrt{} operations.
At a queue depth of 16, for instance, the \fio{} benchmark reveals that 92.35\% of \wrt{} operations were merged (into larger requests).
The \wrt{} throughput saturating at 186\,KIOPS in the inter-zone scenario is reflective of the device performance, as merges are absent in this scenario; the \wrt{} throughput is still higher than the inter- and intra-zone \app{} throughput.
Both \rd{} and \wrt{} operations may also benefit from hardware acceleration, whereas the (first-generation of) \app{} implementations may require extra support from the firmware.
We, hence, expect \app{} performance to improve as ZNS devices mature.
Based on these observations, we recommend intra-zone parallelism for \wrt{} operations (when using \textit{mq-deadline}) for small (i.e., 4\,KiB) I/O requests.

\obs{8}{For large (i.e., $>=$8\,KiB) I/O requests, the performance of intra-zone \app{} and inter-zone \wrt{} operations reach the device limit and scale with concurrency levels in a similar manner.}
We have already seen that request size has a large impact on throughput for requests at a concurrency level of one in \autoref{sec:exp-1-app-write-lat}.
To further investigate the relation between request size and higher concurrency levels, we increase the request buffer size from 4\,KiB to 8\,KiB and 16\,KiB (\autoref{fig:inter-versus-intra-bandwidth}), while retaining the setup intact as in prior experiments.
We issue \app{} operations to a single zone at variable queue depths and \wrt{} operations concurrently (i.e., multiple threads) to multiple zones at a queue depth of 1.
Inter-zone \wrt{} operations initially offer better performance than the intra-zone \app{} operations (the former likely benefits from optimized implementations), they converge quickly to offering similar throughput.
The small (i.e., 4\,KiB) requests fail to reach the device limit ($\sim$1.2\,GiB/s), achieving a maximum throughput of 726.74\,MiB/s for \wrt{} operations, while the large (i.e., $>=$8\,KiB) requests reach the limit when using 2--4 zones concurrently.
Similar to \autoref{sec:exp-1-app-write-lat} we observe that larger requests lead to higher throughput.
\app{} operations scale poorly compared to \wrt{} operations as they require a higher level of concurrency to reach the device limit, and use of multiple zones concurrently benefits \wrt{} more than \app{}.
Although not shown in the figure, when we use high queue depths when we issue \app{} operations to multiple zones, performance degrades:
We observe throughput reductions of up to 20\,MiB/s with a queue depth of 4 for \app{} operations issued to 4 zones concurrently.
We urge use of either intra-zone or inter-zone \app{} operations, but not both.
For bandwidth-intensive workloads (i.e., with request sizes of at least 8\,KiB), we recommend intra-zone \app{} and inter-zone \wrt{} operations.

\rec{2}{Prefer intra-zone to inter-zone parallelism; the former is ideal for \app{} and \rd{} operations, while the latter is best suited for \wrt{} operations. Issue I/O at large request sizes (i.e., $>=$8\,KiB, close to the internal block size), as larger requests scale better with higher concurrency levels.}


\subsection{The Zone State Machine Transition Costs}\label{sec:exp-3-sm}

ZNS offers several unique and novel management operations for interacting with zones, including \open{}, \close{}, \rst{}, and \fns{} (refer~\autoref{sec:background} and \autoref{fig:zns-state-diagram}).
These operations are issued explicitly by an application, but little is known about their performance implications.
We also know little about the performance cost of implicit zone transitions.
Quantifying the cost of these operations is, hence, crucial for designing performance-sensitive applications such as schedulers, file systems, and key-value stores.
We analyze the performance of implicit and explicit operations, the cost of opening and closing zones, and the costs of \fns{} and \rst{} operations.
Since \fio{} does not evaluate or use all relevant state transitions, we evaluate state-transition costs with custom \SPDK{} benchmarks.


\obs{9}{There is no performance difference between explicit and implicit zone open transitions, and the cost of opening/closing is marginal.}
We can open zones either \explicitly{} with the \open{} or \implicitly{} by writing to them (see \autoref{fig:zns-state-diagram}).
We measure the costs of opening a zone under three different configurations:
(1) \textit{explicitly} using an \open{} operation, (2) \implicitly{} with \wrt{}, and (3) \implicitly{} with \app{}.
All \wrt{} and \app{} operations are issued at a request size of 4\,KiB, since we know its latency characteristics (see \autoref{fig:f1-optimal-req-size}).
After opening a zone, we fill it with either \app{} or \wrt{} operations, and we check for any difference in latency performance between \implicitly{} and \explicitly{} opened zones.
We fill the zone to the second-last page, close the zone, and measure the latency of the \close{} operation. 


Our experiments show that it takes about 9.56$\,\mu$s for opening a zone and 11.01\,$\mu$s for closing it.
The costs of the \close{}, \app{}, and \wrt{} operations appear agnostic to how (i.e., \implicitly{} or \explicitly{}) we opened the zone.
The first \wrt{} and \app{} operation to an \implicitly{} opened zone, however, experiences some small (non-trivial) latency overhead---2.02\,$\mu$s and 2.83\,$\mu$s, on average, for \wrt{} and \app{} operations, respectively.
To put in perspective, this is an overhead of about 17.38\% and 19.32\% for 4\,KiB \wrt{} and \app{} operations respectively.
These overheads are not surprising since zones must still be opened before a \wrt{} (or an \app{}) operation can be issued, even if the two operations are merged into one larger operation.
We observe the costs of \open{} and \close{} operations to be marginal.

\begin{figure}[tbp]
  \begin{subfigure}[t]{0.45\linewidth}
    \includegraphics[width=\linewidth]{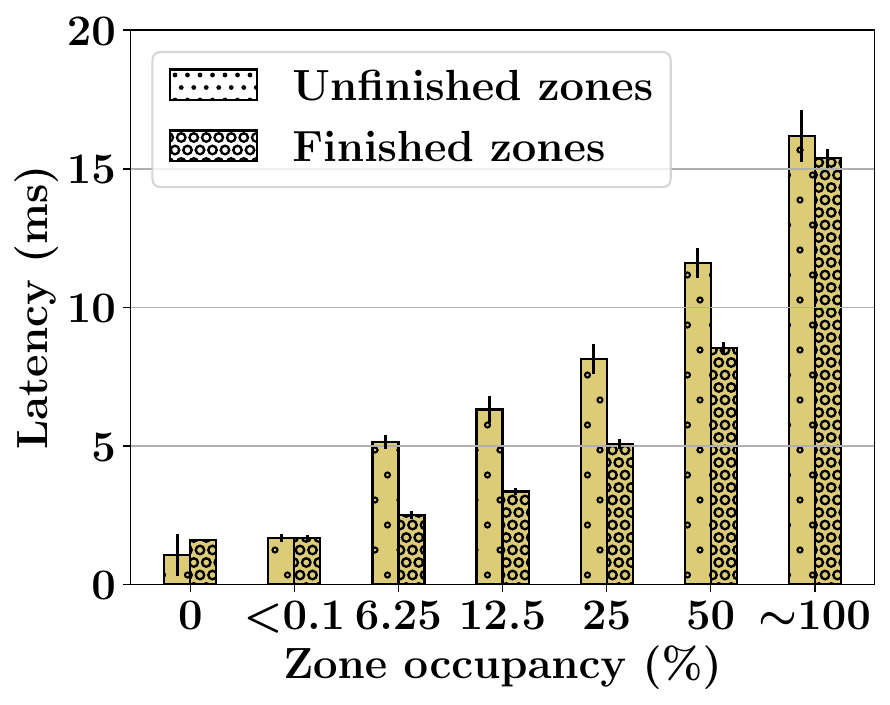}
    \sfigcap{}\label{fig:partial_reset_lat}
  \end{subfigure}
  \begin{subfigure}[t]{0.45\linewidth}
    \includegraphics[width=\linewidth]{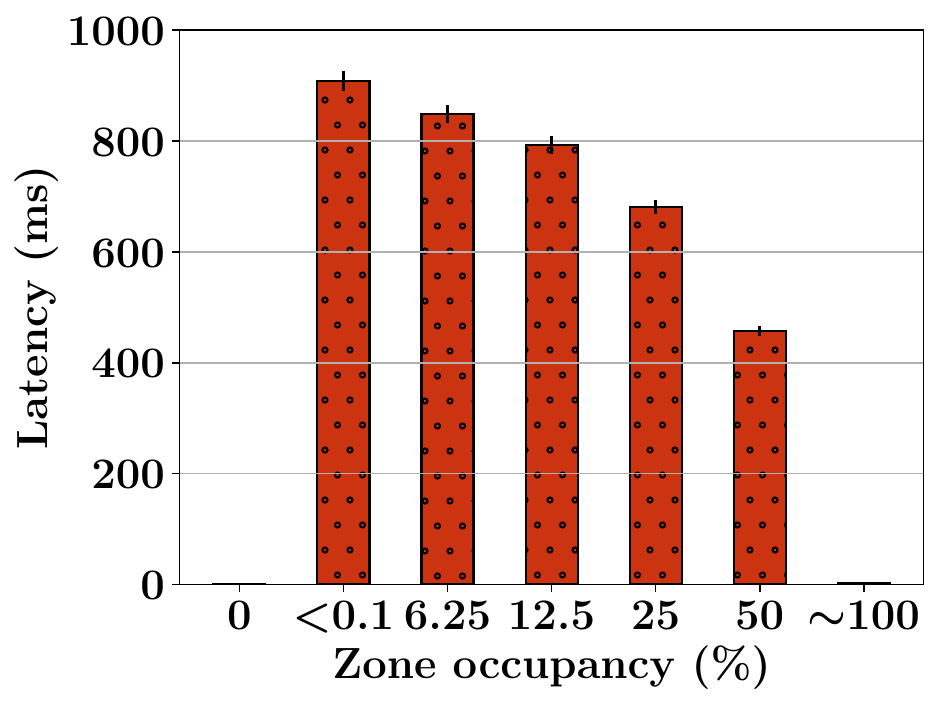}
    \sfigcap{}\label{fig:partial_finish_lat}
  \end{subfigure}
  \figcap{(a) \rst{} latency of partially-occupied zones and (b) \fns{} latency of partially-occupied zones. The y-axis range differs significantly between the plots---up to 20\,ms in (a) and 1,000\,ms in (b).}
\end{figure}


\obs{10}{Zone occupancy (or utilization) has a significant impact on the performance of both \rst{} and \fns{} operations.}
Prior work demonstrates a positive correlation between zone size and \rst{} latency~\cite{jung2023preemptive}, and we ask whether that correlation extends to zone occupancy (i.e., the count of written blocks in a zone) and applies to \fns{} operations as well.
We issue \rst{} operations on 3,000 zones sequentially (over multiple runs) and change the occupancy of zones through various levels---0\% (empty), 1 page (minimal), 6.25\%, 12.5\%, 25\%, 50\% and 100\%.
We use sequential 4\,KiB \wrt{} operations to fill a zone to the desired level.
Once the desired occupancy level is reached, we pause for a second to let the device stabilize as \wrt{} operations influence the \rst{} performance (we defer that discussion until \autoref{sec:exp-5-inf-reset}).
We then issue either a \rst{} operation to the zone, or a \fns{} operation followed by a \rst{} operation.
The two approaches enable us to evaluate the latency of \rst{} operations on both unfinished and finished zones.


Our experiments reveal that zone occupancy has a substantial impact on \rst{}'s performance (\autoref{fig:partial_reset_lat}):
\rst{} operations incur an overhead of 11.60\,ms on half-full zones and 16.19\,ms on full zones, which is three orders of magnitude higher than \wrt{} latencies (see \autoref{sec:exp-1-app-write-lat}).
We posit that \rst{} operations may require metadata updates to unmap the used pages in a zone.
The \textit{trim} operation on conventional NVMe SSDs, which hints to the SSD that it can reclaim a page, also incurs overheads due to metadata updates~\cite{2021-atc-znswap, 2011-sigops-trim}.
We also observe that the latency of \rst{} operations on zones depends on the zone being finished before it was reset.
Finished zones take, for instance, less time to be \rst{} than unfinished zones.
Resetting a half-full zone takes, on average, 3.08\,ms (26.58\%) less time than resetting a zone that was first finished.
Regardless of occupancy, a zone \fns{} operation is, however, a very expensive operation.


We benchmark the performance of the \fns{} operation in a manner similar to how we evaluated the \rst{} operation.
An occupancy of less than 0.1\% (in~\autoref{fig:partial_finish_lat}) indicates that we only filled one page, while $\sim$100\% implies that we filled all except for one; since the standard does not permit us to issue a \fns{} operation to a full or empty zone.
The \fns{} latency, unlike that of \rst{} operations, decreases with occupancy (indeed linearly from $<0.1\%$ to 25\%).
Average latency decreases by about 295 times---from 907.51\,ms (almost one second!) to 3.07\,ms---when we increase occupancy from $<$0.1\% to 100\%.
Hence, when we add the costs of \texttt{finishing} and \texttt{resetting} a zone, the total cost can reach up to hundreds of milliseconds more than that spent in merely \texttt{resetting} the zone.


\rec{3}{Avoid the \fns{} operation (more so than a \rst{}), especially for partially written zones. Minimize the number of zones that need to be finished, hence, by leveraging intra-zone parallelism (thus, reducing the number of active zones, supporting the recommendations in \autoref{sec:exp-2-scalability}).}

\subsection{I/O interference: \wrt{}, \app{} and \rd{} interference}\label{sec:exp-4-inf-io}

\begin{figure}[tbp]
  \begin{subfigure}[t]{0.45\linewidth}
    \includegraphics[width=\linewidth]{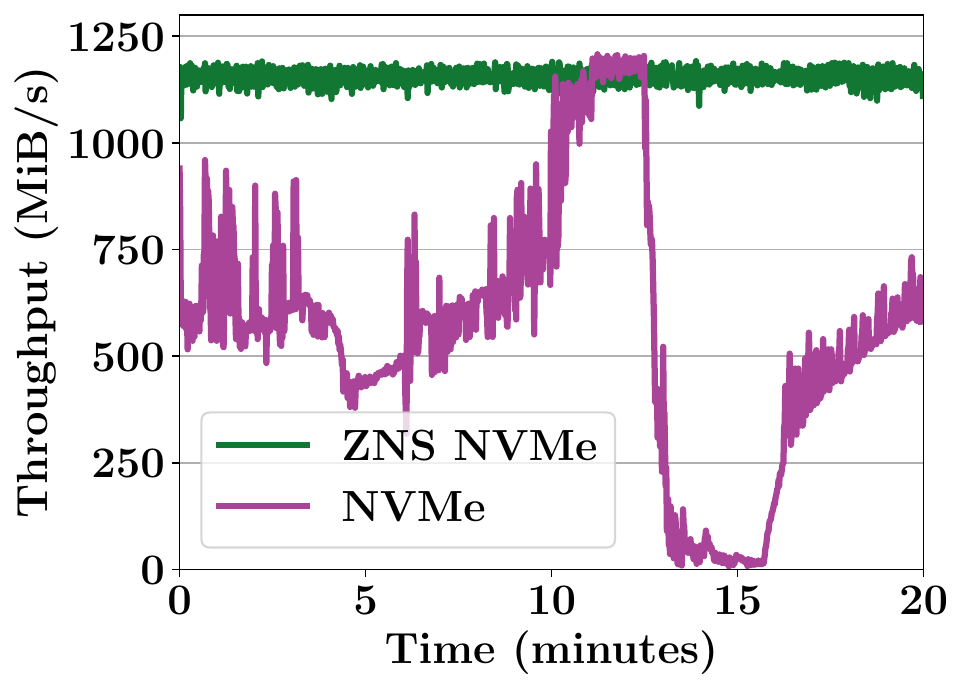}
    \sfigcap{}\label{fig:write_gc}
  \end{subfigure}
  \begin{subfigure}[t]{0.45\linewidth}
    \includegraphics[width=\linewidth]{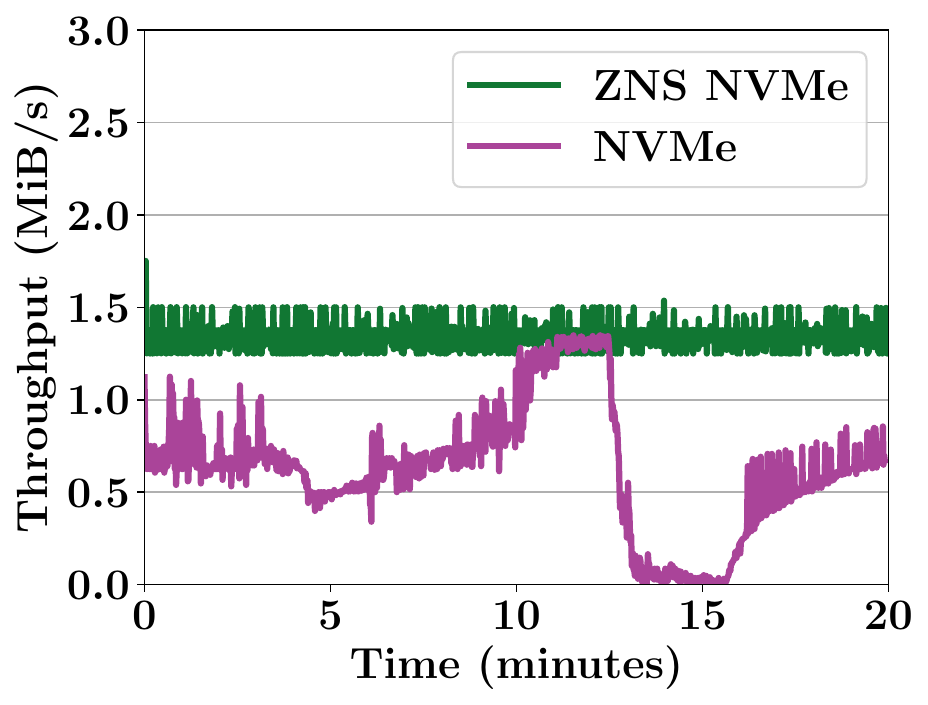}
    \sfigcap{}\label{fig:read_gc}
  \end{subfigure}
  \figcap{(a) \wrt{} throughput of concurrent workloads issuing random \wrt{} operations and (b) random \rd{} throughput while concurrent workloads are issuing random \wrt{} operations. The Y-axis ranges differ substantially between the plots---up to 1,300\,MiB/s in (a) and just up to 3\,MiB/s in (b).}
\end{figure}


Below, we evaluate the interference from write operations (i.e., \wrt{} and \app{}) on read operations (i.e., \rd{}).
Applications typically use workloads characterized by a mix of read and write operations, and it is crucial, hence, to measure the interference caused by read and write operations on each other.
Latencies of \rd{} operations are, for instance, affected significantly by interference from other \rd{} or \wrt{} operations on flash storage~\cite{2017-asplos-reflex}.
Garbage collection (GC) operations happening in the background on conventional SSDs further exacerbate interference, resulting in long \rd{} tail latencies.
GC operations in ZNS, however, are triggered explicitly by the host---a widespread hypothesis is that this approach limits interference effects.
In this section, we investigate this hypothesis.

We design an experiment in which we compare a conventional NVMe SSD with a ZNS SSD, both SSDs have the same hardware specifications.
%
%
%
Fundamental in this evaluation is that GC is triggered inside the FTL firmware on the conventional SSD. 
Triggering GC is fundamental because GC is known to affect the \rd{} throughput negatively.
In ZNS the GC process is managed by the software itself, and thus in this case the benchmark is responsible for the GC.

We evaluate the impact of garbage collection on SSD performance under diverse write workloads, controlled by explicitly rate limiting the workload's write bandwidths.

%
%
We base this limit on the peak ZNS NVMe and ordinary NVMe SSD write bandwidth, which we measured to be 1,155\,MiB/s (see \autoref{fig:inter-versus-intra-bandwidth}).
%
%
The write bandwidth is rate limited to values of 0, 250 (i.e., $\sim$25\%), 750 (i.e., $\sim$75\%) and 1,155\,MiB/s (i.e., 100\%) using \fio{}. 
Concurrently we issue random \rd{} operations and we measure the write and \rd{} throughput over time.
%
%
%
We sent \wrt{}/\app{} operations with four threads; each thread issues 128\,KiB requests at a queue depth of 8.
%
%
We pick this configuration as it puts significant pressure on the SSD, forcing intensive GC.
%
%
We use random \wrt{} operations on the conventional SSD, whereas on ZNS we utilize \app{} operations on a set of random zones.
%
%
Lastly, we issue \rd{} operations randomly at 4\,KiB request size and use one thread (separate from the write threads).
%

\obs{11}{ZNS devices offer more stable read and write performance in the presence of concurrent write-triggered garbage collections than ordinary NVMe devices.}
We observe that ZNS does not have the same write/read throughput fluctuations that we observe on the conventional NVMe interface.
%
%
This observation also confirms earlier ZNS research on throughput stability for ZNS~\cite{2021-atc-zns}.
%
%

We report that both write and read throughput remains stable in all rate-limiting configurations (not shown).
This observation confirms earlier ZNS evaluations~\cite{2020-nvmsa-zns-implications}.
%
%
On conventional SSDs, on the other hand, write and read throughput fluctuates for \textit{all} configurations with concurrent writes.
Especially, when \wrt{} operations are rate limited to the peak bandwidth (i.e., 1,115\,MiB/s).
%
%
We only plot this scenario in \autoref{fig:write_gc} and \autoref{fig:read_gc} for \wrt{} and \rd{} operations, respectively.
%
%
On the x-axis we plot the time in minutes and on the y-axis we plot the throughput in MiB/s.
The results are not surprising as the SSD needs to issue garbage collection (GC) operations in the background during \wrt{}-heavy workloads, which leads to performance drops, while ZNS does not need this (as shown in \autoref{fig:write_gc}).
%
%
\wrt{} throughput fluctuates between a few MiB/s up to 1,200\,MiB/s.
%
%
ZNS also performs GC operations via \rst{}, though the cost of resetting is $\sim$1\% of the cost of filling the zone.
%
%
\autoref{fig:read_gc} shows the impact of GC operations on random \rd{} latency (QD\,32, 4\,KiB). We use QD\,32 as the performance saturates in the experiment at this point. \rd{} latency is affected significantly by concurrent \wrt{} operations and GC operations.
%
%
Furthermore, when \wrt{} operations are rate limited to 1,115\,MiB/s and \rd{} operations are issued at queue depth 1, \rd{} latency at the $95^{th}$ percentile increases to 299.89\,ms for the conventional SSD and 98.04\,ms for the ZNS SSD (not shown in a figure).
%
%
To put it in perspective, when only \rd{} operations are issued, the $95^{th}$ percentile latencies are 81.41\,$\mu$s for both conventional and ZNS, an increment by a factor of 1,000.
%
%

It is important to consider I/O interference when an application needs to scale, even if the interference effects are stable.
For example, the observations made in \autoref{sec:exp-2-scalability} do not consider concurrent I/O. 
Achievable throughput becomes limited if there is concurrent I/O. Intra-zone scalablity will, therefore, saturate at lower queue depth and inter-zone scalability at fewer concurrent zones. 
Coincidentally, an application that uses high intra- or inter-zone scalability reduces the throughput of other applications.
Applications must hence take interference effects into account to achieve QoS targets.
In short, inter- and intra-zone resources are shared between multiple applications/threads.

\rec{4}{Developers should measure the peak read/write performance of ZNS devices, and provision their application storage needs around them. There is no need to account for performance fluctuations because of GC operations.}
%
%

\subsection{Reset interference}\label{sec:exp-5-inf-reset}

\begin{figure}[t]
  \centering
  \includegraphics[width=0.55\linewidth]{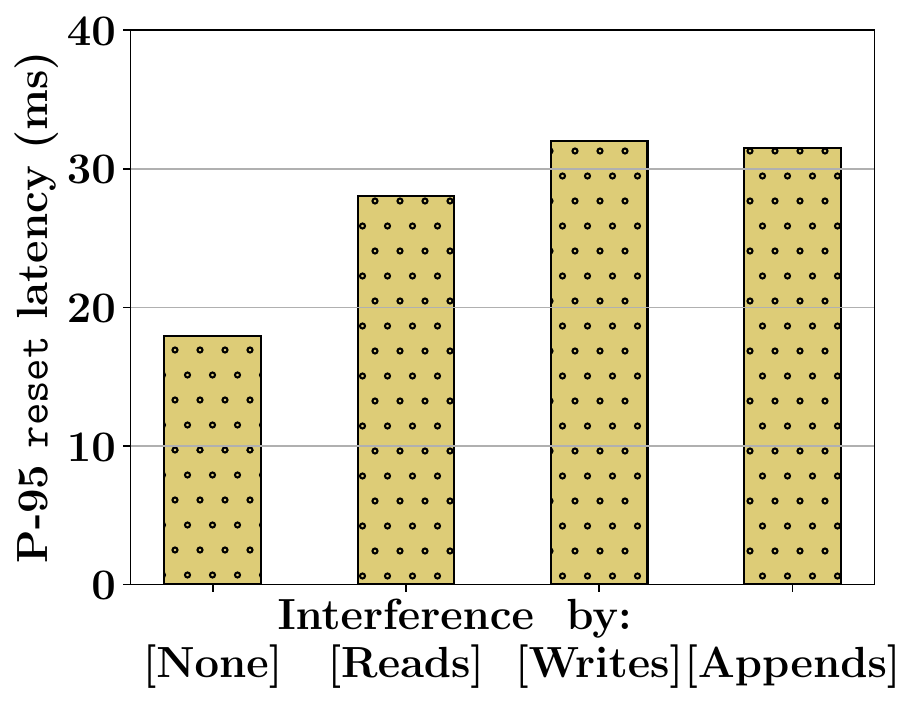}
  \figcap{Interference effect of \rd{}, \wrt{}, and \app{} operations on the 95-percentile \rst{} latency}
  \label{fig:reset_inteference}
\end{figure}

We observe that the \rst{} operations take tens of milliseconds to complete (\autoref{sec:exp-3-sm}), \textit{provided that we execute them in isolation}.
%
%
In real workloads or applications, however, it is likely that we issue \rst{} operations concurrently with other I/O, e.g., as a separate garbage collection thread, and not in isolation.
%
%
In this section we evaluate the effects of concurrent I/O on \rst{} operations and the other way around. 
This evaluation is also important to support the observations made in \autoref{sec:exp-4-inf-io}.
We evaluate with a custom benchmark using \SPDK{}. 
%
%

In this benchmark we use two concurrent threads.
We use one thread solely for issuing \rst{} operations on 100\% occupied zones and one for issuing either \app{}, \rd{}, or \wrt{} operations.
%
%
The intent is to ensure that there always is concurrent I/O and to measure if the latency alters during concurrent load.
%
%
To prevent \rst{} and \rd{} operations or writes to the same addresses, we issue \rst{} operations to the first half ($\sim$400 zones) of the device and \rd{}, \wrt{} and \app{} operations to the second half.
%
%
We issue \wrt{} and \app{} operations sequentially, but \rd{} operations randomly.
%
%
We set the request size to 4\,KiB.
%
%
We finish the experiment once there are no more zones to \rst{}, and repeat the experiment three times.
%
%
%

\obs{12}{\rst{} operations do not interfere with \rd{}, \wrt{} or \app{} latency.}
%
%
We discover no significant effect of \rst{} operations on either \app{}, \wrt{} or \rd{}.
%
%
Such behavior is likely to happen when the \rst{} operation is limited to metadata operations on different resources than are used by I/O.
Note that such behavior can be device-specific, but does give the impression that there is little need for schedulers to account for \rst{} when managing QoS for \rd{}, \wrt{}, or \app{}.

\obs{13}{\rd{}, \wrt{} and \app{} operations interfere with the \rst{} latency significantly.}
%
%
Contrary to the previous observation, we do observe that \app{}, \wrt{} and \rd{} operations influence \rst{} latency.
%
%
In \autoref{fig:reset_inteference} we plot this interference effect.
On the X-axis we plot the operation that runs concurrently with \rst{} operations (if any), and on the Y-axis we plot the $95^{th}$ percentile tail latency of \rst{} operations.
%
%
We can infer that the $95^{th}$ percentile \rst{} latency increases from 17.94\,ms to 28.00\,ms for concurrent \rd{} operations~(56.11\% increase), 32.00\,ms for concurrent \wrt{} operations~(78.42\% increase) and 31.48\,ms for concurrent \app{} operations~(75.50\% increase). All three I/O operations thus increase \rst{} latency significantly. Therefore, it is likely that there is a form of resource contention, where I/O operations are prioritized over \rst{} operations inside of the ZNS device.

\rec{5}{\rst{} operations can be issued concurrently with I/O operations like \rd{}, \wrt{} and \app{}, since \rst{} operations do not have an impact on I/O latency. Further on, while \rst{} latency itself does increase significantly when they are issued concurrently with other I/O operations, \rst{} operations themselves are issued at a large granularity (per-zone, 1 GiB), leading to sporadic \rst{} operations (minimum time between two \rst{} operations is: zone\_size / write\_bandwidth, on our device about one second).
}

\section{Open challenges with ZNS emulation}

So far we reported numbers for a specific device, the \href{https://www.westerndigital.com/products/internal-drives/data-center-drives/ultrastar-dc-zn540-nvme-ssd#0TS2094}{ZN540}. 
However, the actual ZNS design/implementation space is large. 
Here, NVMe emulators can help: They have been used in previous research~\cite{2022-hotstorage-comp-lsm, liu2022fair, oh2021efficient, 2021-osdi-zns+}. 
The use of emulators can help verify observed trends and generalize our results. 
We observe, however, that none of the available emulators are ZNS-ready (yet). 
Emulators that are currently publicly available for ZNS include FEMU~\cite{2018-fast-femu} and NVMeVirt~\cite{kim2023nvmevirt}. 
The more recent ConfZNS~\cite{song2023confzns}, which we discuss only briefly, was not an open-source software at the time of our evaluations.
We only consider the design/implementation of the emulators in this section and do not re-run the evaluations for the emulators:
As of this writing, we could not run all of our evaluations in the emulators due to stablility or compatibility issues in the emulators (e.g., \textit{SPDK} not working).
We assume that as the software ecosystem stabilizes, these issues will be resolved.
The goal of this section is, therefore, to address challenges in the design of emulators themselves. 

We explain which of our ZNS performance observations the available emulators FEMU and NVMeVirt are currently unable to capture because of their design and explain what should be changed to support them. 
%
%
We do not consider observations \#1 (LBA format), \#2 (\SPDK{} has the lowest I/O latencies), and \#11 (ZNS is more stable than NVMe) to be relevant in this section as they do not represent essential behavior to emulate (e.g., not ZNS-specific). 

FEMU currently makes no attempt at emulating ZNS SSD request latency, and requests are, therefore, as fast as the underlying hardware (i.e., CPU and DRAM) permits. 
The lack of a latency model leads to various reproducibility challenges.
First, it is not possible to reproduce differences in I/O latency behavior of \app{}, \wrt{}, and \rd{} operations, since there is no inherent performance difference between these operations. 
There is also no difference in intra- and inter-zone scaling.
As a result, FEMU fails to reproduce real-world empirical observations \#3, \#4, \#5, \#6, \#7, and \#8. Similarly, FEMU does not emulate zone transition latency, and as a result FEMU fails to also reproduce \#9, \#10, \#12, and \#13.
For example, \fns{} operations will become unrealistically fast as they are limited to metadata operations in DRAM.
Therefore, FEMU, in its current state, cannot accurately reproduce {\em any} of our real-world observations and, consequently, cannot be used for evaluating the performance of real ZNS applications. 
We recommend FEMU to take an approach similar to NVMeVirt, as it has an explicit timing model. 
This timing model accounts for channel and NAND latency, and distinguishes between \rd{} and \wrt{} latency.

NVMeVirt utilizes a latency model that is shown to be reasonably accurate for ZNS devices~\cite{kim2023nvmevirt}. 
However, currently NVMeVirt uses the same latency model for both \app{} and \wrt{} operations and, hence, cannot represent observation \#4 (\app{} and \wrt{} latency differ). 
This shortcoming introduces similar issues for intra- and inter-zone scaling, as \app{} and \wrt{} latencies are equal. 
This issue would prevent  \#5 and  \#6 to be accurate. We argue that NVMeVirt should use a different model for both operations.
For example, by using a different latency for \app{} operations.
Unfortunately, NVMeVirt also does not represent zone management operations correctly. 
It sets the latency of \rst{} static and equal to NAND erasure latency (multiple milliseconds), which we have observed to not always be the case. Therefore, we opt that it should also be possible to use a dynamic cost where the latency depends on zone occupancy. 
A simple linear model would suffice here. Notably, NVMeVirt does not emulate timing for the other zone management operations at all. 
Since the \fns{} operation is shown to be expensive, we argue that this operation should be implemented, preferably with a model based on zone occupancy. 
As it stands NVMeVirt fails to reproduce \#9, \#10, \#12, and \#13. 
In short, while NVMeVirt is accurate for \rd{} and \wrt{} operations, it requires more refinement for \app{} and zone transition operations to be a more accurate model.

More recently, Song et al. released ConfZNS, a ZNS emulator with an accurate latency model~\cite{song2023confzns}.
%
%
As of this writing, the code was not open-source; we do not, hence, describe the accuracy of this emulator in-depth and are unable to investigate which observations can be reproduced.
The timing model of ConfZNS promises to lead to accurate latencies of \wrt{} operations for inter-zone and \rd{} operations for inter- and intra-zone scalability, which should lead to results similar to \autoref{sec:exp-2-scalability} and is the first step towards designing accurate ZNS emulators.
%

%

In short, while available open-source emulators have latency models for \wrt{} and \rd{} operations, no current emulator has (or proposes) an accurate latency model for either \app{} operations or zone transitions. 
Emulators should consider adopting both in order to be accurate.


\section{Related Work}\label{sec:rwork}

\subsection{On Zoned Namespace (ZNS) devices} 
There have been a number of (performance) characterizations for ZNS, but none consider the \app{} operation, the ZNS state transitions and the interference of ZNS operations. 
Nevertheless, existing ZNS evaluations/implementations on various other properties, such as zone isolation, do exist and are complementary to this work. 
One of the early performance studies is by Shin et al.~\cite{2020-nvmsa-zns-implications}, where they verified the performance isolation properties of ZNS devices, and show that increasing I/O size decreases I/O latency, confirming our claims on interference and request size. Similar to our study, they investigated inter-zone scalability and the impact of request sizes.
Bae et al.~ investigate the impact that zone size and ZNS internal parallelism have on the host I/O performance~\cite{2022-hotstorage-zns-parallelism}. 
They identify that large zone sizes are preferred as they offer more opportunities to stripe and distribute I/O requests across multiple parallel channels and flash dies.
We show that this is true for \app{}s as well.
However, large zones (in GiB/s) are also said to have large zone \rst{} latencies that can significantly influence \rd{} latencies (pushing them to milliseconds and seconds). 
In our results we have only observed \rst{} latencies in the milliseconds, but also showed that the cost is not static and largely depends on zone occupancy. 
We could not reproduce the effects of \rst{}s on \rd{} latency. 
As a result of the aforementioned \rst{} cost, Bae et al. advocate using small zones~\cite{2022-hotstorage-zns-parallelism}.  
To improve the device performance and parallelism, the authors introduce a host-side inference tool to identify zone parallelism mappings by inter-zone interference measurements, and an accompanying I/O scheduler that can do mapping-aware I/O scheduling. 
Im et al.~  use small-zone SSD inter-zone parallelism for RocksDB~\cite{2022-middleware-rocksdb-smallzone}. 
Their results on inter-zone parallelism confirm our results and show similar trends for both small and large zone SSDs. 
We limit our evaluation to a large zone SSD in this evaluation.
Jung et al.~ investigate various \rst{} algorithms and show a correlation between zone size and \rst{} latency \cite{jung2023preemptive}. 
Their work complements our findings on zone occupancy.

There is also a healthy amount of research on the ZNS specification and its new operations~\cite{2021-atc-zns,2022-cidr-append,2021-osdi-zns+,2021-hotstorage-rocks}. 
Bj{\o}rling et al.~ present a comprehensive work on ZNS devices itself and discuss the design rationale and integration options~\cite{2021-atc-zns}.
The work also evaluates ZNS on the macro level, while we evaluate ZNS on the micro level.
ZNS's unique \app{} operation is discussed here~\cite{2020-vault-append}. 
Purandare et al.~  discuss the impact of ZNS devices on log-based data management systems, specifically log-based file systems, key-value stores (LSM tree), and database systems with logs~\cite{2022-cidr-append}. 
They identify the ZNS \app{} operation as a unique operation to leverage in the design of these systems. 
Despite much enthusiasm regarding the ZNS \app{} operation, to the best of our knowledge, it has only been used in a handful of systems such as TropoDB~\cite{2022-tropodb}, BtrFS~\cite{rodeh2013btrfs} and ZNSwap~\cite{2021-atc-znswap}. 

There is also ample research in the application domain. 
%
%
Especially on the KV-store RocksDB and the ZenFS file system-backend is prevalent~\cite{2022-hotstorage-comp-lsm,2022-zenfs,2022-hotstorage-llc, 2022-hotstorage-comp-lsm, 2020-hotstorage-zns-lsm-gc, 2022-zenfs,2022-middleware-rocksdb-smallzone, 2021-atc-zns}. 
ZNS application research has, as of now, mostly focused on improving garbage collection algorithms and new zone allocation policies. 
Such work can now also use the observations made in this device characterization. 
For example, by accounting for \fns{} latency and zone occupancy for their garbage collection algorithms.

Naturally, this large body of work identifies that there is a big interest in accommodating and adapting ZNS-capable storage devices in the storage software stack.
In all of these previous works, there has been a selective performance benchmarking and characterization using a mix of real-device~\cite{2021-wd-zns}, emulators~\cite{2018-fast-femu, kim2023nvmevirt}, or hardware support~\cite{2021-osdi-zns+}.
In this work, we primarily focus on aspects of performance of a commercially available ZNS device. 
In this process we have verified past published results as well as and reported new results. 
%
 
\subsection{On Performance Characterization}
Due to the black-box nature of flash SSDs, multiple past studies have focused on empirical, stochastic, and analytical modeling of their operational characteristics~\cite{2021-tos-ssd-blackbox,2018-ieee-ssd-bottleneck-analysis,2010-mascots-ssd-blackbox-modeling,2011-mascots-ssd-modeling}.
Such modeling is important for accurately predicting the access latency of a flash SSD for performance provisioning and QoS-oriented scheduling.
Past works have extended HDD-based blackbox analytical models to flash SSDs (e.g., linear, or regression based) at a broader workload-level granularity~\cite{2010-mascots-ssd-blackbox-modeling,2011-mascots-ssd-modeling}. 
SSDCheck studies the impact of write buffering and garbage collection by implementing various representative algorithms in SSD hardware, and verify/map their results to multiple blackbox SSDs~\cite{2021-tos-ssd-blackbox}.
In comparison, our work is focused on measurement-based study to establish the baseline performance without deconstructing the ZNS internals.  
    
The modeling and impact of garbage collection algorithms have been one of the most studied areas with flash SSDs~\cite{2018-tos-ibm-flashsystems,2019-tom-d-choice-gc,2021-sigmetric-ssd-management,2012-sigmetrics-ssd-perf-anamoly,2013-sigmetrics-gc-modeling, 2015-sigmetrics-ssd-optimal-greedy}.
Hu et al.~ analytically model the residual lifetime of a given SSD, in the presence of GC and enterprise workload~\cite{2012-sigmetrics-ssd-perf-anamoly}.
Pletka et al.~ present the details of enterprise-grade latency, ECC, and GC flash algorithms~\cite{2018-tos-ibm-flashsystems}.
Li et al.~ present a stochastic Markov chain model to model the I/O dynamics of an SSD with concurrent I/O and GC workloads~\cite{2013-sigmetrics-ssd-stochastic-modeling}. 
Using this model, they design a randomized greedy algorithm that can be tuned to operate close to the optimal operational curve of the SSD. 
As write amplification, wear-leveling and the choice of GC algorithm are inextricably linked, Verschoren and Houdt analyze (e.g., simulation, and trace-based workloads) d-choice GC algorithms for their impact on flash wear-leveling and device lifetime~\cite{2019-tom-d-choice-gc,2013-sigmetrics-gc-modeling}.
Lange, Naor and Yadgar take a broad approach to SSD performance modeling and consider the ``SSD management'' problem that includes block allocation, wear leveling, write amplification, and garbage collection from an algorithmic perspective~\cite{2021-sigmetric-ssd-management}.  
They report on a series of analytical models which are verified with synthetic and trace-based workloads.

Before the emergence of ZNS devices, there has been work on modeling SMR device operations~\cite{2015-tos-skylight}. 
Shafaei et al.~ present an analytical model for a device-managed SMR drive (unlike ZNS which is host managed)~\cite{2017-tos-device-mng-smr,2016-sigmetrics-smr-short}.
Chen et al.~ present one of the earliest systematic studies focused on flash SSD performance characterization~\cite{2009-sigmetrics-flash-ssd-study}. 
They identified various undocumented performance anomalies due to flash fragmentation and establish a high correlation between access pattern and flash performance (previously thought to be unrelated).
Jung and Kandemir provide a thorough and detailed empirical evaluation of six SSDs for their \rd{}, \wrt{}, and \texttt{trim} (similar to the ZNS \rst{} operation) interference from background activities (GC and buffer flush) performances~\cite{2013-sigmetrics-ssd-expectations}.
Their results show unexpected influences of \rd{}s on the device lifetime (P/E cycle) and the influence of background activities on sustained SSD performance. 
In their work, they also recommend exposing flash firmware API to the host software, that ZNS does in a standard way.    
Our work follows the same spirit for a new generation of flash storage devices with unique operations. 
 

     
\section{Threat to Validity}
Much of our results in this work confirm the expected ZNS behavior that is hypothesized in ZNS's development. The confirmation of these results also opens up new directions of research where much of the previously published work on flash SSD is open for scrutiny---and perhaps can even become obsolete~\cite{2021-hotos-zone}. 
Our experiments and benchmarks are \textit{empirically driven, user-observed} behavior for one specific type of ZNS device. Such selective benchmarking has risks. 
\textbf{We have consulted and verified our observations with Western Digital, the ZNS device manufacturer, for the particular ZNS device that we have tested.} However, we are aware that it is challenging to generalize our findings as many device internal details are confidential, and ZNS device capabilities are expected to improve in the future. 
Nonetheless, we believe that this paper makes strong contributions by performing a first-of-its-kind systematic performance sweep of a NVMe Flash Device with Zoned Namespaces, and providing specific workable 
recommendations to the developers. In order to ensure the long-term viability and repeatability of our research we have open-sourced all the scripts, tools, and data sets collected. We are also in the process of acquiring different ZNS models to extend our study.  

\section{Conclusion}

Zoned Namespace-capable NVMe devices represent a significant step in the evolution of flash hardware and software stack.
They offer a rich interface (introducing operations such as \app{}, \fns{}, and \rst{}) to the host software that allows fine-grained control over managing the flash storage.
In this work, we systematically characterized the performance of a commercially available ZNS-capable NVMe device.
To this end, we developed benchmarks and tools to characterize the performance the new I/O operation (i.e., \app{}) and flash-management commands (i.e., \fns{}, \rst{}, \texttt{open}, and \texttt{close}).
We analyzed the impact of operation interference (I/O and management) between conventional and ZNS-capable NVMe devices.
We present five recommendations to ZNS application developers concerning \app{} performance, inter-zone and intra-zone scalability, the cost of management operations, and I/O- and GC-level interference.
We identify shortcomings in the state-of-art ZNS emulators, which are widely used in academic research, and outline the changes that they require to ensure a high fidelity emulation.
We published all the artifacts of this study at \url{https://github.com/stonet-research/NVMeBenchmarks}.
We hope our results and the publicly available artifacts encourage developers and researchers to apply and evaluate our recommendations in a wide variety of applications and expand this work with similar characterizations of other ZNS devices.



\section*{acknowledgment}

This work is supported by a generous donation of NVMe (ZNS) SSDs from Western Digital and the Dutch Research Council (NWO) grant number OCENW.KLEIN.561.
Krijn Doekemeijer is funded by the VU PhD innovation program.
We want to thank the anonymous reviewers for their invaluable feedback and the AtLarge team from the Vrije Universiteit Amsterdam for their continued support.
%
%

\bibliographystyle{ieeetr}
\bibliography{main}

\appendix

\begin{figure*}[tbp]
  \centering
  \begin{subfigure}[t]{0.3\linewidth}
    \includegraphics[width=\linewidth]{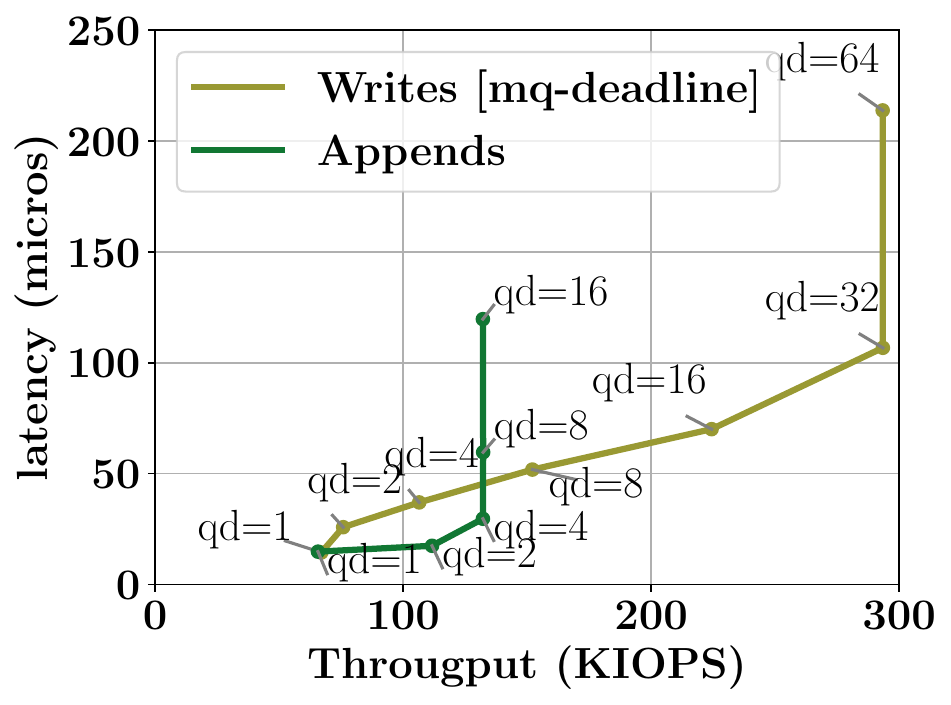}
    \sfigcap{}\label{fig:kneeplot_4k}
  \end{subfigure}
  \begin{subfigure}[t]{0.3\linewidth}
    \includegraphics[width=\linewidth]{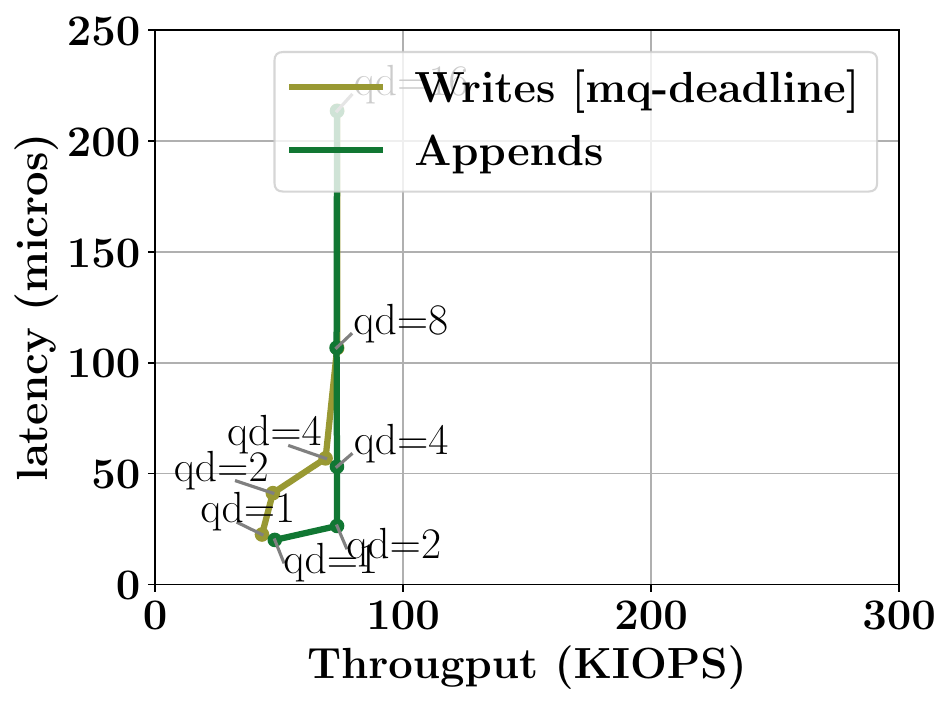}
    \sfigcap{}\label{fig:kneeplot_16k}
  \end{subfigure}
  \begin{subfigure}[t]{0.3\linewidth}
    \includegraphics[width=\linewidth]{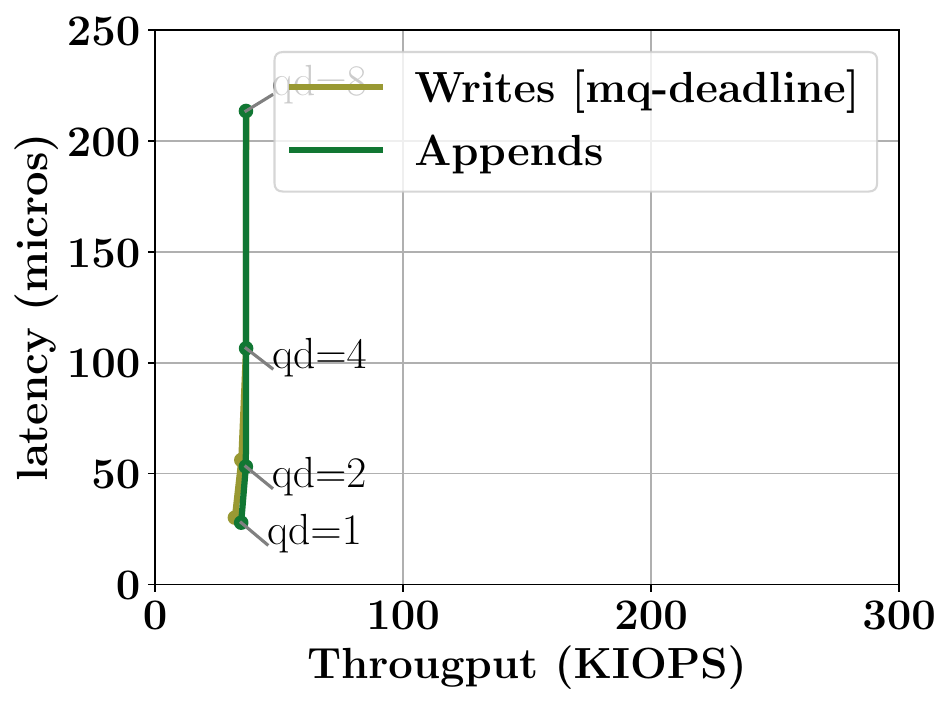}
    \sfigcap{}\label{fig:kneeplot_32k}
  \end{subfigure}
  \figcap{\app{} and \wrt{} throughput/latency at various queue depths: (a) 4KiB Requests; (b) 16KiB Requests; and (c) 32KiB Requests; concurrency level is queue depth for \app{} operations, and concurrent zones for \wrt{}s.}
  \label{fig:kneeplot_append_write}
\end{figure*}

There is one additional result we did not present in our main contributions as it's impact is lower.
This result is related to request latency at higher queue depths for \app{} and \wrt{} operations, a scenario which will be common in real-life workloads. 
We chose to include it in the Appendix to aid further research.

In \autoref{fig:kneeplot_append_write} we plot the effect of higher queue depth for both \app{} and \wrt{} latencies. 
In the plot, on the x-axis the throughput (higher is better) is plotted and  on the y-axis the request latency (lower is better).
The experiment setup as the same as used for \autoref{fig:intra_throughput}. \wrt{} operations are send with \iou{} to a single zone and use the \mq{} scheduler, and \app{} operations are send with \SPDK{}.
We observe that as the queue depth increases, both the latency and throughput increase.
However, the latency of \wrt{} operations increases significantly more than \app{} operations until a certain threshold.
Past this threshold (4 for all block sizes), the latency trends are the same.
From these results, we can recommend two things: (1) \app{} operations should only be send at low queue depth to get the best latency; (2) intra-zone scalability with \app{} operations is preferred over \wrt{} operations as it leads to lower latency.


%
%
%
%
%
%
%

\end{document}